\documentclass[twoside,english]{elsarticle}
\usepackage[utf8]{inputenc}
\usepackage[T1]{fontenc}
\usepackage{geometry}
\usepackage{dsfont}
\usepackage{color} 
\usepackage{amsmath}
\usepackage{amssymb}
\usepackage{graphicx}
\usepackage{nicematrix}
\usepackage{cancel}
\usepackage[textsize=scriptsize, backgroundcolor=yellow!10, linecolor=gray!35, textwidth=2.5in]{todonotes}

\makeatletter
\journal{\vspace*{1cm}}

\usepackage{babel}
\usepackage{hyperref}
\usepackage[normalem]{ulem} 

\usepackage{array}
\usepackage{booktabs}
\usepackage{multirow}
\usepackage{rotating}

\makeatother

\begin{document}

\begin{frontmatter}{}

\title{Initiation of motility on a compliant substrate}

\author[LiPhy]{Jocelyn \'Etienne}

\ead{jocelyn.etienne@univ-grenoble-alpes.fr}

\author[LiPhy]{Pierre Recho\corref{cor1}}

\ead{pierre.recho@univ-grenoble-alpes.fr}

\cortext[cor1]{Corresponding author}

\address[LiPhy]{Université Grenoble Alpes, CNRS, LIPHY, 38000 Grenoble, France}

\begin{abstract}
The conditions under which biological cells switch from a static to a motile state are fundamental to the understanding of many healthy and pathological processes. In this paper, \textcolor{black}{we consider a cell constrained to move along a one-dimensional track.} We show that even in the presence of a fully symmetric protrusive activity at the cell edges, such a spontaneous transition can result \textcolor{black}{from a feedback of the deformation of an elastic substrate on the cell traction forces.} The loss of symmetry of the traction forces leading to the cell propulsion is rooted in the fact that the surface loading follows the substrate deformation, \textcolor{black}{leading the cell to surf its own wake}. The bifurcation between the static and motile states is characterized analytically and, considering the measurements performed on two cell types, we show that such an instability can realistically occur on soft \emph{in vivo} substrates. 
\end{abstract}

\end{frontmatter}{}

\section{Introduction}

Most living cells have the ability to move to perform collective tasks necessary to vital biological functions such as wound healing or the immune response. To do so, they rely on their cytoskeleton, an active biopolymer meshwork that consumes chemical energy to locally contract and turnover \cite{boal_2012}. Simple continuum models have been used to physically understand how either non-uniform contractility driven by molecular motors or material turnover due to the cytoskeleton polymerization and depolymerization can drive cell crawling by producing asymmetric traction forces on the cell substrate  \cite{Julicher+Joanny.2007.1,Lin2010model,carlsson2011mechanisms,bergert2015force, Recho-Truskinovsky.2015.1,Drozdowski+Schwarz.2021.1, chelly2022cell}. Moreover, starting from a static state, cells on stiff substrates can spontaneously initiate their motion by breaking the symmetry of their locomotion apparatus \cite{verkhovsky1999self,yam2007actin,barnhart2015balance}, following for instance a chemomechanical instability forming a pattern \cite{mori2008wave,copos2020hybrid,lavi2020motility}, dynamically modulating the adhesion \cite{Sens.2020.1} or the polymerization \cite{Ron+Gov.2020.1}, or through a material instability whereby the active stress created by some molecular motors self-amplifies the cell cytoskeleton flow \cite{recho2013contraction, tjhung2012spontaneous,callan2013active, giomi2014spontaneous}. It has also been shown that a cell formed of an isotropic gel, with constant surface polymerization and bulk depolymerization, can break its initially symmetric configuration through a morphological instability when the protrusive forces overcome the membrane resistance \cite{blanch2013spontaneous}. However, such an instability requires a two-dimensional shape and has no analogue for cells constrained to move along a narrow track, \textcolor{black}{either at the surface of a substrate \cite{maiuri2012first,Mohammed+Gabriele.2019.1} or  inside a capillary \cite{hawkins2009pushing}  which is the framework considered in this paper. Interestingly, it has been argued that motility confined to such 1D structures  may be closer to the physiological 3D motility  in Extra-Cellular Matrices (ECM) than 2D motility on a flat substrate \cite{doyle2009one,doyle2013dimensions}.} 

In this paper, we show that substrate deformation can lead to symmetry breaking through a mechanical feedback that initiates  the motion of a cell with symmetric boundary polymerization and bulk depolymerization. 
In many existing models of mechanotaxis, the observed influence of the substrate stiffness or deformation on cell motility \cite{lo2000cell,duchez2019durotaxis} is included through various mechanosensing pathways controlling the cell adhesion \cite{lelidis2013interaction,lober2014modeling,feng2019cell,sens2013rigidity,chen2022unified}, the molecular motors' active stress \cite{banerjee2011substrate,shenoy2016chemo} or the cytoskeleton polymerization rate \cite{zhang2020minimal,oliveri2021theory}.  \textcolor{black}{In  contrast, rather than mechanosensing, it is here the action and reaction at the cell--substrate interface  which is at the origin of the spontaneous cell polarization. Indeed, the coupling of
the internal cell dynamics and the substrate deformation is shown to result in a `follower load' instability \cite{bigoni2011experimental}.}

The paper is structured as follows. In Sec.~\ref{sec:model}, we introduce the mechanical model describing the flow of the cell cytoskeleton based on material turnover and its frictional interaction with the substrate. We then show in Sec.~\ref{sec:ext_act} that the treadmilling motion of the cytoskeleton makes it possible to transport the cell with an externally imposed substrate deformation. Considering in Sec.~\ref{sec:coupled_pb} a semi-infinite incompressible elastic substrate under small deformations, which are now due to the traction forces of the cell itself, we show in Sec.~\ref{sec:lin_fric} that when the cell-generated protrusive forces exceed a finite threshold, the static and symmetric state becomes unstable to the benefit of a polarized motile state. This instability occurs within the range of realistic material parameters values but features a substrate deformation field that is not admissible, as it implies matter self-penetration. It is then shown in Sec.~\ref{sec:nl_friction} that this  degenerate situation can be eliminated without compromising the instability  when considering a biologically relevant non-linear friction case. Finally in Sec.~\ref{sec:large_def}, using a simpler Winkler-type model where the substrate consists of an elastic film anchored to a stiff foundation, we analyse the influence of large deformations which automatically prevent matter self-penetration. We show again that this does not qualitatively compromise the cell polarization instability described in Sec.~\ref{sec:lin_fric}. Interestingly, it is also shown that non-linear substrate deformations can lead to a reentrant behavior whereby the cell can spontaneously polarize and then depolarize as the symmetric protrusive forces increase.

\section{Model of the cell skeleton: treadmilling-driven retrograde flow}\label{sec:model}

We consider an active gel slab \cite{kruse2005generic,kruse2006contractility}
 moving along a one-dimensional track, modeling the biopolymer meshwork of a cell crawling on a fibronectin-coated track. 
Neglecting inertial effects, the momentum balance in the gel reads 
\begin{equation}\label{e:momentum_bal}
h\partial_x\sigma=f,
\end{equation}
where $h$ is the constant gel height, $\sigma(x,t)$ is the axial stress in the gel and $f(x,t)$ is the traction force  applied by the cell on the substrate. The spatial coordinate of material points in the gel at time $t$ is denoted $x\in[l_-(t),l_+(t)]$, where $l_-(t)$ and $l_+(t)$ are the moving fronts of the gel. We denote by $C(t)=(l_-(t)+l_+(t))/2$ the segment center and by $L(t)=l_+(t)-l_-(t)$ its length.
In the absence of any external loading at the segment ends, the stress vanishes at the cell boundaries,
\begin{equation}\label{e:momentum_bal_bc}
\sigma(l_{\pm}(t),t)=0.
\end{equation}



\begin{figure}[t]
\hfill{}\includegraphics [scale=0.9]{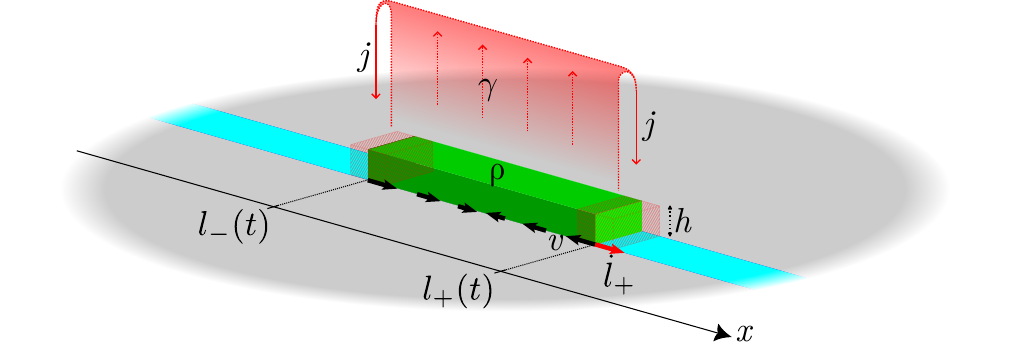}\hfill{}
\caption{\textcolor{black}{Schematic of the active gel segment on a track (light blue) subjected to a symmetric polymerization flux balanced by bulk depolymerization.}}
\label{f:profiles}
\end {figure}

In order to clarify the physical ingredients at play in the instability that we shall present, we voluntarily neglect the active stress due to the presence of molecular motors and consider only the activity due to the meshwork polymerization at the boundaries and its depolymerization in the bulk.
Mass balance within the gel reads \cite{kruse2005generic}
\begin{equation}\label{e:conservation_mass}
\dot{\rho}+\rho \partial_x v=-\gamma \rho, 
\end{equation}
where $\rho(x,t)$ is the mass density of the gel and $\gamma$ is a bulk depolymerization rate. 
The superimposed dot $\dot{}=\partial_t+v\partial_x$ denotes the total time derivative. This bulk mass conservation equation is associated with the boundary conditions:
\begin{equation}\label{e:conservation_mass_bc}
\rho(l_{\pm}(t),t)(v(l_{\pm}(t),t)-\dot{l}_{\pm})=\pm j,
\end{equation}
where $j<0$ is a fixed and \emph{symmetric} polymerization flux at the two cell ends. See schematic in Fig.~\ref{f:profiles}.

The constitutive behavior of the gel is modeled as an elastic solid at a short timescale. Hence, $\sigma$  is a increasing function of the strain $\rho_*/\rho$  that vanishes when $\rho=\rho_*$, the reference density of the  biopolymer meshwork.  See \cite{putelat2018mechanical} for further discussion on the effective mechanical properties that arise in these conditions. \textcolor{black}{As we explain below, due to our assumption that the relative density variations are small (see \eqref{e:tech_assumption}), the specific dependence of the stress on the density of the gel is irrelevant in our analysis.}

Combining the polymerization fluxes \eqref{e:conservation_mass_bc}  with the absence of applied stress at the boundaries  \eqref{e:momentum_bal_bc} \textcolor{black}{which imposes $\rho=\rho_*$ there,} \textcolor{black}{we obtain  the widely used kinematic conditions for the moving fronts \cite{larripa2006transport,kruse2006contractility,recho2013asymmetry, blanch2013spontaneous,Ambrosi-Zanzottera.2016.1}:
\begin{equation}\label{e:conservation_mass_bc_bis}
\dot{l}_{\pm}=\pm v_p+v(l_{\pm}(t),t),
\end{equation}
where $v_p=-j/\rho_*>0$ is a \emph{symmetric} polymerization velocity at both ends \cite{carlsson2011mechanisms,Recho-Truskinovsky.2015.1,Drozdowski+Schwarz.2021.1}.}

Next, we assume that both the magnitude and rate of fluctuations of the gel density are small: 
\begin{equation}\label{e:tech_assumption}
|\rho-\rho_*|/\rho_*\ll 1\quad\text{and}\quad  \dot{\rho}/\rho\ll \gamma
\end{equation}  
 In this case, similarly to \cite{blanch2013spontaneous}, \eqref{e:conservation_mass} leads to $\partial_xv=-\gamma$. The cytoskeleton flow in the cell is then linear:
 $v(x,t)=-\gamma (x-l_-(t))+\bar{v}(t)$, with $\bar{v}(t)$ a time-dependent function to be determined. This expression of velocity, combined with \eqref{e:conservation_mass_bc_bis}  leads to the following fronts dynamics: $\dot{L}=-\gamma L+2v_p$ and $\dot{C}=\bar{v}-\gamma L/2$, showing that $L$ converges to a steady state fixed value $L=L_{\text{eq}} = 2v_p/\gamma$ and that $\bar{v}$ can be expressed as a function of $\dot{C}$:  
 $$v(x,t)=-\gamma (x-C(t))+\dot{C}(t).$$

Such a symmetricaly treadmilling system can be put into motion by a spatial asymmetry in its frictional interaction with the substrate \cite{Sens.2020.1,Ron+Gov.2020.1,schreiber2021adhesion} since the traction forces are then biased in one direction. We consider here a different situation where the friction coefficient $\xi$ with the substrate is a constant but the traction forces depend on the relative velocity of the gel with respect to the substrate: 
\begin{equation}\label{e:traction}
  f = \xi\Psi_*(v- v_s). 
\end{equation}
In \eqref{e:traction},  $v_s$ is the tangential velocity of the substrate surface along the track and $\Psi_*$ is an odd function, which characterizes the non-linear behavior of the sliding friction between the gel and the substrate. This is relevant to cell migration,
since the effective friction law is known to be biphasic \cite{gardel2008traction} due to the stochastic collective dynamics of force-sensitive adhesive bonds \cite{sabass2010modeling,sens2013rigidity}. In what follows, we take the simple approximation 
$$\Psi_*(v) = \frac{v}{1+(v/v_*)^2},$$
which leads to  a switch of behavior when the relative velocity reaches the threshold velocity $v_*$. 

We aim at showing  that the non-local feedback of the friction forces through the velocity $v_s$ of a compliant substrate can lead to a symmetry breaking and spontaneous motility.

\section{Externally actuated substrate}\label{sec:ext_act}
To first explain the physics of this motion, we start by considering the situation of a deformation imposed externally to a linear elastic incompressible substrate, which occupies the half space $z<0$  \textcolor{black}{below the cell}. We assume that the deformation is created by the presence of magnetic beads at the substrate surface, along the track. The beads are actuated by an electromagnet of exactly the cell size $L$ that is quasistatically moved along the track at a velocity $V_e$. As they are polarized in the $z$ direction, the torque acting on the beads vanishes while they experience a traction force oriented only in the $x$-direction that takes the form $f_m=f_m^0F(y(x,t))$, where the characteristic force $f_m^0$ scales with the current in the electromagnet and the beads magnetization and $y(x,t)=2(x-V_e t)/L$ is the traveling wave coordinate. By asymptotic matching of the magnetic field \textcolor{black}{(see \cite{Landau+.1984.8}, chap IV, paragraph 30)} close and far from the magnet, we can approximate $F(y)=-y \text{ if } |y|<1 \text{ and } -y/|y|^{5} \text{ if } |y|>1 $. Next, restricting our analysis to small deformations, the substrate displacement at the surface $z=0$ is purely tangential and takes the form of the traveling wave  $u(x,t)=u^0U(y(x,t))$ where, in plane strain, $U$ reads \textcolor{black}{(see \cite{Johnson1987} paragraph 2.4 for the derivation of this logarithmic plane strain kernel)}:
\begin{equation}\label{e:substrat_vel_0}
U(y)=-\int_{-\infty}^{\infty}\log |y-y'|F(y') \mathrm{d}y', 
\end{equation}
and $u^0=3f_m^0L/(2\pi E_s)$. The Young modulus of the substrate is denoted $E_s$.

\begin {figure}[t]
\hfill{}\includegraphics[scale=0.7]{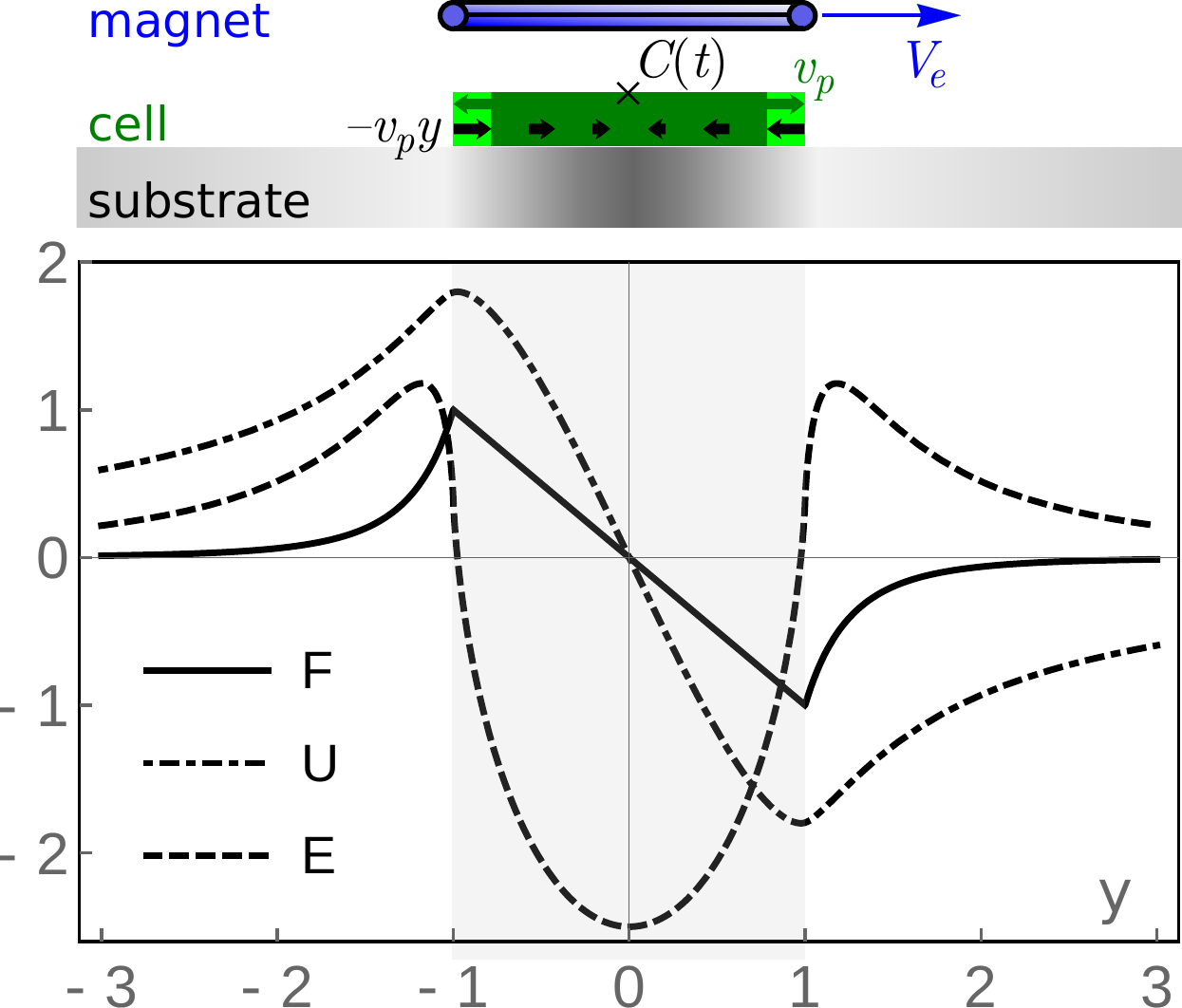}\hfill{}
\caption{Top, an electromagnet moved at velocity $V_e$ induces a mechanical strain (gray-coded) in a substrate. A cell is in frictional interaction with the substrate, it is assumed to treadmill with velocity $-v_py$ and polymerize at its edges with speed $v_p$.
Bottom, shape of the force applied by the magnet, displacement and strain of the elastic substrate surface.}
\label{f:cell_driven_magnetic}
\end {figure}

In \eqref{e:substrat_vel_0}, we have neglected the contribution of the cell to the traction force acting on the substrate, assuming that $|f_m^0|\gg\xi v_p$. Using again the small deformation assumption, the substrate velocity is also a traveling wave given by $v_s=\partial_tu=-V_e\epsilon$ where the strain of the substrate surface is $\epsilon=\epsilon^0E(y(x,t))$ with $E(y)=\partial_yU$ and $\epsilon^0=3f_m^0/(\pi E_s)$.  The functions $F$, $U$ and $E$ representing the spatial variations of the applied force, substrate displacement and strain along the track are shown in Fig.~\ref{f:cell_driven_magnetic}.

 Finally, as the total force that is exerted on the treadmilling segment has to vanish, we obtain from \eqref{e:traction}  that 
\begin{align}
&0= -\int_{C(t)-L/2}^{C(t)+L/2} f(x) \mathrm{d}x,\text{ which implies that, }0= \int_{-1}^{1} \Psi_*\left(\dot{C}(t)-v_py+V_e\epsilon^0E\left(y+\delta(t)\right)\right) \mathrm{d}y\label{e:surfer},
\end{align}
where $\delta(t)=y(C(t),t)$. Equation \eqref{e:surfer} is a differential equation that sets the $C(t)$ dynamics. Taking $C(0)=0$ (the driving electromagnet is centered with the cell from the start) and since $E$ decays to zero at infinity, the only permanent regime of motion is for $\delta(t)=0$, i.e.\ a segment velocity equal to the actuation velocity $\dot{C}=V_e$: we call this mode of motion at the speed of the elastic wave `surfing'.
Thus, \eqref{e:surfer} becomes the  algebraic equation 
\begin{equation}
0= \int_{-1}^{1} \Psi_*(-v_py+V_e(1+\epsilon^0E(y))) \mathrm{d}y,
\end{equation}
that sets the strain magnitude $\epsilon^0$ of the traveling wave such that it can be surfed by the treadmilling gel.

If $\Psi_*$ is a linear function, this surfing condition becomes independent of $V_e$ and $v_p$ and simply reads $\bar{\epsilon}=\frac{1}{2}\int_{-1}^{1}\epsilon(y)\mathrm{d}y=-1$. Such an average elastic strain of the substrate is not admissible since it implies self-interpenetration of matter. Likewise, for non-linear $\Psi_*$, no solution exists if the treadmilling is slow, $v_p \ll V_e(1-\epsilon^0)$. We thus focus on the case $v_p\gg V_e$. Since $\int_{-1}^{1} \Psi_{*}(-v_py)\mathrm{d}y=0$, a first order expansion  leads to the surfing condition
\begin{equation}
0= \int_{-1}^{1} \Psi_{*}'(-v_py)\left(1+\epsilon(y)\right) \mathrm{d}y.
\label{e:surfing_condition}
\end{equation}
 The above condition, where~${}'$~denotes the derivative, shows the importance of the treadmilling dynamics and the non-linear friction for the surfing  to be physically feasible. Indeed, the prefactor $\Psi_{*}'(-v_py)$ has to change sign in order to make the cancellation of the integral possible for admissible $\bar{\epsilon}>-1$. The necessary and sufficient conditions for surfing motion are thus the presence of treadmilling and a biphasic friction $\Psi_*$. These conditions are in fact necessary  whatever the origin of the substrate deformation.
 
\section{The coupled problem}\label{sec:coupled_pb}

We now turn to the coupled problem asking whether the cell can surf on an elastic wave that it itself creates instead of relying on an external traveling wave driven by a moving magnetic field. We therefore express the surface displacement in response to the cell traction forces themselves \cite{Johnson1987}, which we had previously neglected compared to the action of the magnetic field:
\begin{equation}\label{e:substrat_vel}
u(x,t)=\frac{-3}{2\pi E_s}\int_{C(t)-L/2}^{C(t)+L/2}\!\!\log\left| \frac{x-x'}{L}\right|f(x'\!,t) \mathrm{d}x'\!. 
\end{equation}
Changing the time and space variables to follow the cell motion, $s=t$ and $y=2(x-C(t))/L$, we have $\partial_t=\partial_s-(2\dot{C}/L)\partial_y$. Then, the velocity of the substrate for small deformations $v_s=\partial_tu$ can be expressed as a functional of the traction force distribution from \eqref{e:substrat_vel}:
\begin{align}\label{e:v_s}
v_s(y,s)&=-\frac{3L}{4\pi E_s}\int_{-1}^{1}\log|y-y'|\partial_sf(y'\!,s) \mathrm{d}y'+\frac{3 \dot{C}(s)}{2\pi E_s}\int_{-1}^{1} \frac{f(y'\!,s)}{y-y'} \mathrm{d}y'\!.
\end{align}
In these coordinates, the cell cytoskeleton's velocity is
\begin{equation}\label{e:v}
v(y,s)=-v_py+\dot{C}(s).
\end{equation}
Non-dimensionalizing the distance by $L$, the time by $1/\gamma$ and the stress by $2\xi v_p$, and defining the non-dimensional substrate displacement and strain ($\epsilon=\partial_xu$):
\begin{equation}
 u[f](y,s)=-\frac{\theta}{2}\int_{-1}^{1}\log|y-y'|f(y'\!,s) \mathrm{d}y' \quad\text{and}\quad
\epsilon[f](y,s)=-\theta\int_{-1}^{1}\frac{f(y'\!,s)}{y-y'} \mathrm{d}y'\!,
\end{equation} 
the three relations $\eqref{e:traction}$, \eqref{e:v_s} and \eqref{e:v} are combined in a single non-linear integral equation that determines the distribution of the cell traction forces, 
\begin{equation}\label{e:integ_pb}
2\beta f=\Psi\left(2\beta(\dot{C}-y/2 -u[\partial_sf]+\dot{C}\epsilon[f])\right),
\end{equation}
with $\Psi(v) = v/(1+v^2)$.
The non-dimensional parameters of the model, respectively 
$$\beta=\frac{v_p}{v_*}\quad\text{and}\quad\theta =\frac{3\xi v_p}{\pi E_s},$$
represent the level of importance of the non-linearity of the sliding friction law ($\beta\ll 1$ being the linear viscous friction limit, \textcolor{black}{either because $v_*$ is large or because $v_p$ is small)} and the stress imposed by the protrusion compared to the elastic modulus of the substrate. \textcolor{black}{Rough estimates of these model parameters, characteristic scales and non-dimensional parameters in the well studied case of fish keratocytes are compiled in Table~\ref{t:valpar}}. 
\begin{table}
\begin{center}
\color{black}
\begin{tabular}{lll}
\hline\hline
name & symbol & typical value  \\ 
\hline
actin depolymerization rate & $\gamma$ & $0.03\text{ s}^{-1}$  \citep{laurent2005gradient,rubinstein2009actin}\\
actin polymerization velocity & $v_p$  &$0.25$ $\mu\text{m~s}^{-1}$  \cite{rubinstein2009actin} \\
friction coefficient & $\xi$ &$1$ kPa~s~$\mu\text{m}^{-1}$  \citep{rubinstein2009actin,barnhart2011adhesion} \\
saturation velocity & $v_*$ &$ 0.125$ $\mu\text{m~s}^{-1}$ \cite{barnhart2015balance} \\
ECM Young modulus & $E_s$  &$100$ Pa \citep{levental2009matrix, dolega2021mechanical}\\
\hline
characteristic length & $L=2v_p/\gamma$ &$17$ $\mu$m  \\
characteristic time & $1/\gamma$ &  $33$ s \\
characteristic velocity & $L\gamma$ & $0.5$ $\mu\text{m}\,\text{s}^{-1}$ \\
characteristic stress & $2\xi v_p$ &  $0.5$ kPa \\
\hline
protrusive stress/substrate stiffness & $\theta=3\xi v_p/(\pi E_s)$ & 2.5\\
traction forces saturation  & $\beta=v_p/v_*$ & 2 \\
\hline\hline
\end{tabular}
\end{center}
\caption{\textcolor{black}{Rough estimates of the model parameters, characteristic scales and dimensionless parameters for fish keratocytes.\label{t:valpar}}}
\end{table}
Equation \eqref{e:integ_pb}  also depends on the unknown crawling velocity $\dot{C}$, which is set by the global force balance condition
\begin{equation}\label{e:integ_cond}
\int_{-1}^{1}f(y,s)\mathrm{d}y=0.
\end{equation}

When the substrate is infinitely rigid ($\theta=0$), our model \eqref{e:integ_pb} directly provides the expression of the traction forces  $f=\Psi(2\beta(\dot{C}-y/2))/(2\beta)$ and \eqref{e:integ_cond} necessarily implies that $\dot{C}=0$. Therefore, as can be expected for a system with symmetric polymerization at both ends, in the absence of intra-cellular feedbacks of the type considered in \cite{Ron+Gov.2020.1,Sens.2020.1} (a polarization of the adhesion molecules concentration in \cite{Sens.2020.1} and an inhibitor of polymerization in \cite{Ron+Gov.2020.1}), the steady state traction force profile $f_0(y)=\Psi(-\beta y)/(2\beta)$ is always symmetric with respect to the cell center and the cell always remains static. While such a distribution  remains a solution of the integral problem \eqref{e:integ_pb}-\eqref{e:integ_cond}, 
\textcolor{black}{one can note the presence of the nonlinear term $\dot{C} \epsilon[f]$ in
\eqref{e:integ_pb}. This term finds its origin in the time evolution of the substrate deformation due to the dynamics $\dot{C}$ of the interval over which the load $f$ is applied on the substrate, and thus of the motion of the cell itself. We now proceed to examine the stability of this system.}

\section{Linear friction}\label{sec:lin_fric}

To do so, we begin by investigating the situation when $\beta\ll 1$ (the sliding friction law is linear) and \eqref{e:integ_pb} reduces to the linear integral  equation \textcolor{black}{on the distribution $f$:}
\begin{equation}\label{e:integ_pb_lin}
 f=\dot{C}-y/2 -u[\partial_sf]+\dot{C}\epsilon[f].
\end{equation}
\textcolor{black}{Notice however that \eqref{e:integ_pb_lin} still contains an implicit non-linearity in its last term as $\dot{C}$ is a functional of $f$ fixed by \eqref{e:integ_cond}. We show below that the presence of this crucial term, which advects the substrate deformation with the cell motion, is the root of spontaneous motility in this system.}

Inserting in \eqref{e:integ_pb_lin} the perturbation of the static solution with a growth rate $\lambda$,
$$f(y,s)=f_0(y)+\eta\text{e}^{\lambda s} \delta f(y) 
\quad\text{and}\quad
\dot{C}=\eta\text{e}^{\lambda s} \delta\dot{C},$$
we find, at first order in the small parameter $\eta$, that
$\delta f(y)=\delta\dot{C}(1+\epsilon[f_0(y)])-\lambda u[\delta f(y)]$. Imposing the global force balance condition \eqref{e:integ_cond}, $\delta\dot{C}$ is expressed as a functional of $\delta f$  to obtain,
\begin{align}\label{e:perturb_lin} 
&\left( 2+\int_{-1}^1\epsilon[f_0]\mathrm{d}y\right) \frac{\delta f}{\lambda}=(1+\epsilon[f_0])\int_{-1}^1u[\delta f]\mathrm{d}y -\left( 2+\int_{-1}^1\epsilon[f_0]\mathrm{d}y\right) u[\delta f].
 \end{align}
 When $\theta$ is small, multiplying  \eqref{e:perturb_lin} by $\delta f$ and integrating, leads to
$$\int_{-1}^{1}(\delta f(y))^2 \mathrm{d}y=\frac{\lambda\theta}{2}\int_{-1}^{1}\int_{-1}^{1} \log|y-y'|\delta f(y)\delta f(y')\mathrm{d}y\mathrm{d}y'.$$
As the linear operator $u[f]$ is symmetric positive definite \cite{estrada1989integral}, we conclude from the above equation that $\lambda$ is negative and the static steady state is stable to small perturbations for $\theta\ll 1$. This limiting behavior also shows that, in general, the $u$ term in \eqref{e:integ_pb_lin} describing the elastic feedback of the substrate when the loading domain is fixed does not destabilize the central symmetry of the traction forces in the static configuration and thus does not contribute to making the system motile. However, the growth rate of the instability can vanish ($\lambda=0$)  when $\theta$ reaches a critical threshold $\theta_c$ in \eqref{e:perturb_lin} that satisfies the condition $2+\int_{-1}^1\epsilon[f_0]\mathrm{d}y=0$, leading to $\theta_c=2$. Close to this threshold, the behavior of $\lambda(\theta)$ can be expanded in power series and we find at first order that $\lambda\sim 9 (\theta -\theta_c)/(2(\pi^2-6))$, showing that the growth rate of a perturbation of the static solution becomes positive beyond $\theta=\theta_c$.	Thus, the contribution of the substrate deformation due to the fact that the loading domain moves if the cell moves  (so-called `follower load' effect \cite{bigoni2011experimental}) makes the static state unstable when the traction forces compared to the substrate elasticity increase beyond the critical threshold $\theta_c$.
\begin {figure}[!t]
\hfill{}\includegraphics[scale=0.8]{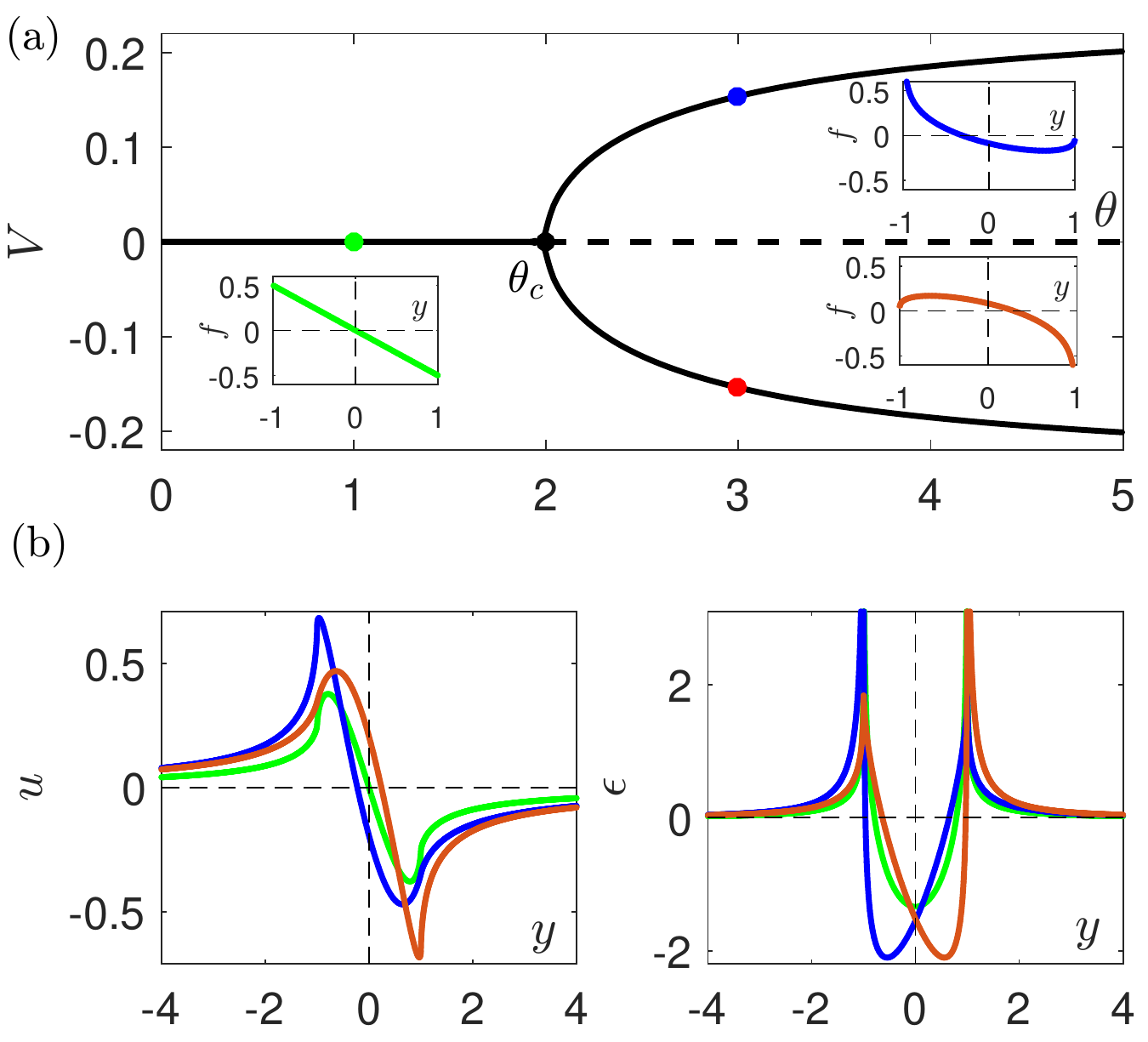}\hfill{}
\caption{ (a) Static to motile transition for a linear sliding friction law (Eq.~\eqref{e:integral_eq_simpl_nd_TW}). Characteristic traction forces profiles \eqref{e:analytical_f} are shown in inserts. The static branch is dashed passed the critical point $\theta=\theta_c$ since it becomes unstable. (b) Profiles of the substrate displacement and strain corresponding to the traction forces shown in (a)}
\label{f:pitchfork}
\end {figure} 

The post-bifurcation regime can be  characterized  analytically by searching traveling wave solutions of \eqref{e:integ_pb_lin} for which $\partial_s f=0$ and $V=\dot{C}$ is a constant.  Indeed, the Cauchy integral equation:
\begin{equation}\label{e:integral_eq_simpl_nd_TW}
\theta V\int_{-1}^{1}\frac{f(y')}{y-y'} \mathrm{d}y'+f(y)=-\frac{y}{2}+V,
\end{equation}
has an integrable solution which reads \cite{Karpenko1966approximate}:
\begin{equation}\label{e:analytical_f}
f_{\text{eq}}(y)=-\frac{(y/2+V)w_{a}(y)}{\sqrt{1+(\theta \pi V)^2}},
\end{equation}
where $w_{a}(y)=((1-y)/(1+y))^a$ and $a=\text{arctan}(\theta \pi V)/\pi$. Thus, imposing the global force balance condition \eqref{e:integ_cond}, we find that the cell velocity satisfies the implicit relation:
\begin{equation}\label{e:velocity_eq}
V=a/2=\text{arctan}(\theta \pi V)/(2\pi).
\end{equation}
If $\theta\leq \theta_c= 2$, equation \eqref{e:velocity_eq} has the single root $V=0$ (stable static solution) while if $\theta>\theta_c$, it has three roots: $V=0$ corresponding to the now unstable static configuration and the two others corresponding to two symmetric motile configurations, see Fig.~\ref{f:pitchfork}(a).

\begin {figure}[!t]
\hfill{}\includegraphics [scale=0.7]{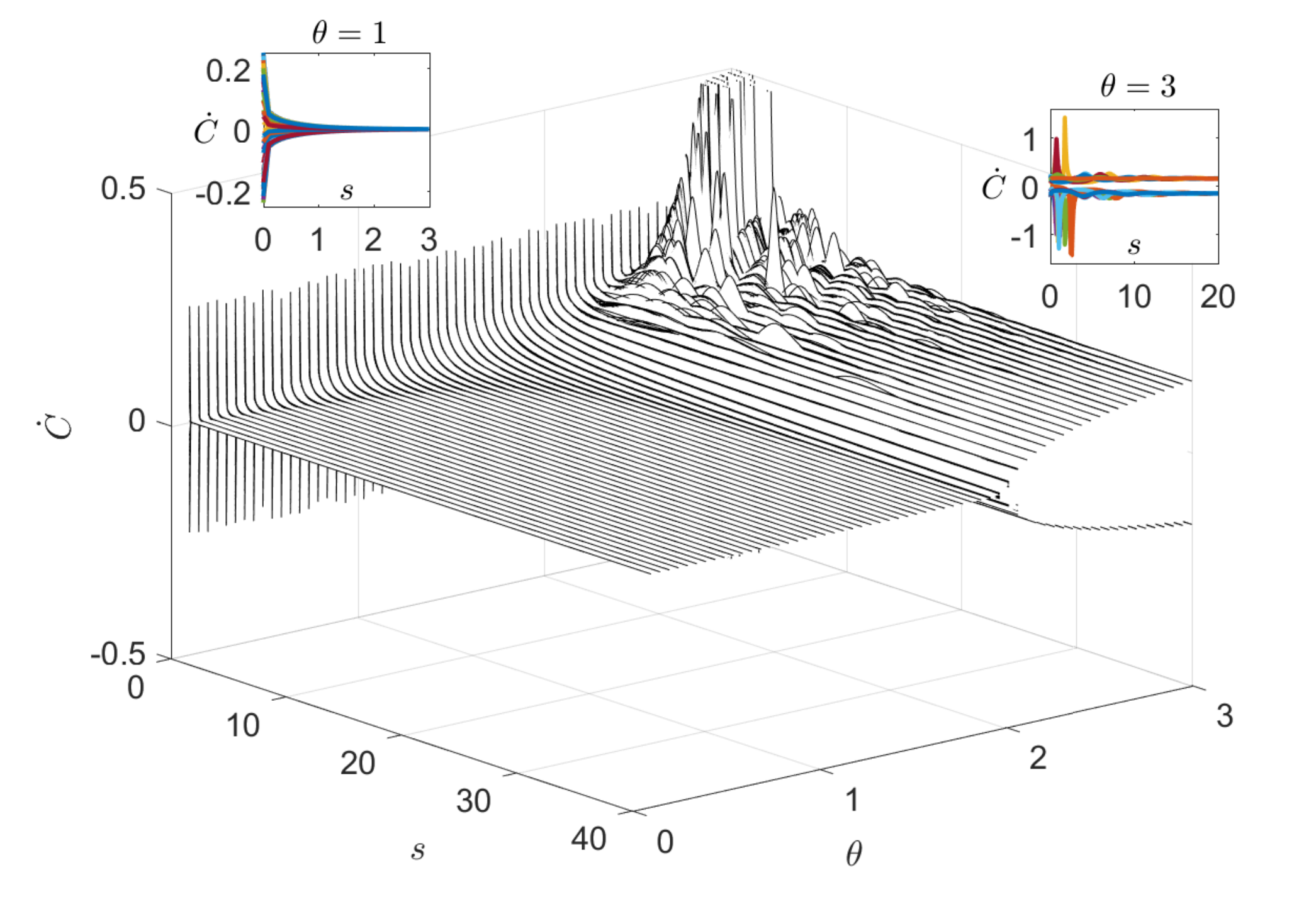}\hfill{}
\caption{Convergence of the instantaneous cell velocity found from \eqref{e:integ_pb_lin} to its steady state value predicted by \eqref{e:velocity_eq} for an initial traction force profile of the form of \eqref{e:analytical_f} but with an initial velocity randomly chosen in the interval [-0.25,0.25] (corresponding to the maximal velocity in \eqref{e:velocity_eq}). Large transient oscillations can arise before reaching the steady state velocity.}
\label{f:stability}
\end {figure} 
\textcolor{black}{Using the model parameters estimates for fish keratocytes collected in Table~\ref{t:valpar}, we find} a value of $\theta\simeq 2.5$, close to $\theta_c$, for a substrate 
with a Young modulus of $E_s\simeq100 \text{ Pa}$\textcolor{black}{, which is within the range of stiffnesses of  ECM measured \emph{ex-vivo} \cite{levental2009matrix} and \emph{in vitro} \cite{dolega2021mechanical}}. This suggests that the present motility initiation instability could realistically be at play following a uniform actin polymerization activation at the membrane for \emph{in vivo} soft substrates. Upon the bifurcation, the symmetry of the traction forces is lost (insets of Fig.~\ref{f:pitchfork}(a)) with a maximum at the trailing edge  and a negative part at the leading edge in qualitative agreement with experiments~\cite{hennig2020stick}. This in turn leads to an asymmetric substrate displacement and strain (see Fig.~\ref{f:pitchfork}(b)) that promotes the  cell motion, which `surfs' on the deformation it creates. 
\textcolor{black}{Conceptually, this can be likened to the ``walker droplets'' that are made to bounce on a liquid bath \cite{Protiere+Couder.2006.1} with an external actuation. Such bouncing generates a surface wave and, beyond a threshold of forcing,  the droplet starts to propagate at a constant velocity due to its interaction with the wave it produces. Yet another example of spontaneous polarization where, contrary to our framework, inertial forces play an important role is that of a bristle-bot robot driven by an oscillating force \cite{desimone2012crawling, cicconofri2015motility}. In this case, it is the dynamical dependence of the friction coefficient itself on the normal component of the traction force that leads to directional motility through a stick-slip instability.}
By simulating numerically \eqref{e:integ_pb_lin}, we  show in Fig.~\ref{f:stability} the robust convergence of the cell to the aforementioned steady state configurations regardless of initial conditions, in line with the small perturbation stability results.

\section{Non-linear friction}\label{sec:nl_friction}

Although the above results with a linear friction law limit ($\beta\ll 1$) allow to characterize the physics of substrate deformation-induced motility initiation, the instability arises in a parameter range for which the deformation is not physically admissible. Indeed, as in the case of externally imposed deformations that has been presented in Sec.~\ref{sec:ext_act}, the surface strain at the bifurcation threshold 
$\epsilon_c(y)=2 (y \log(\sqrt{\left| (1+y)/(1-y)\right| })-1)$ has values below $-1$ over a finite interval. 
We now show that here too, a biphasic sliding friction (finite $\beta$) restores the possibility of an admissible substrate deformation.
In this case, the perturbation of the static state leads to
\begin{equation}\label{e:perturb}
\delta f(y)=\Psi'(-\beta y)(\delta\dot{C}(1+\epsilon[f_0(y)])-\lambda u[\delta f(y)])
\end{equation}
and the growth rate of the perturbation switches from a negative to a positive value at the critical value: 
\begin{equation}\label{e:theta_c_frict_nl}
\theta_c(\beta)=\frac{4 \left(\beta ^2+1\right)}{\left(\beta^{-2}+1\right) \text{arctan}(\beta ) \left(\left(\beta ^2+1\right) \text{arctan}(\beta )+2 \beta \right)-1}.
\end{equation}
The strain at the bifurcation threshold reads
$$\epsilon_c=-\theta_c(\beta)\frac{\beta  y \log \left| \frac{1-y}{1+y}\right| +2 \text{arctan}(\beta )}{2 \left(\beta +\beta ^3 y^2\right)},$$
which approaches zero when $\beta$ increases except in narrow boundary layers at $y=\pm 1$ close to the cell fronts. This leads to an average strain under the gel
$\bar{\epsilon}_c=\frac{1}{2}\int_{-1}^1\epsilon_c(y)\mathrm{d}y = -\theta_c(\beta) \text{arctan}^2(\beta)/(2\beta^2)$,
that tends to zero as $\beta\rightarrow \infty$. Thus for a large enough $\beta$, the critical value of $\theta$ at which the bifurcation happens predicted by \eqref{e:theta_c_frict_nl} becomes small and the small deformations assumption is also verified. For fish keratocytes, the value of $v_*\simeq 0.125 \,\mu\text{m s}^{-1}$ in the biphasic relation was estimated in \cite{barnhart2015balance} based on experiments on stiff substrates, leading to $\beta\simeq 2$ and $\bar{\epsilon}_c\simeq -0.25$. 
The biphasic relation was also initially characterized for another cell type (PtK1). In this case, the polymerization velocity can be estimated from \cite{ponti2004two} to be $v_p\simeq 0.03\,\mu\text{m.s}^{-1}$ and the data of \cite{gardel2008traction} indicates an effective friction coefficient of $\xi\simeq 10\text{ kPa.s.}\mu\text{m}^{-1}$ and a threshold velocity $v_*\simeq 0.01 \,\mu\text{m.s}^{-1}$. This leads to non-dimensional parameters $\theta\simeq 3 $ and $\beta\simeq 3$ 
and to $\bar{\epsilon}_c\simeq -0.15$. 

\begin {figure}[!t]
\hfill{}\includegraphics[scale=0.7]{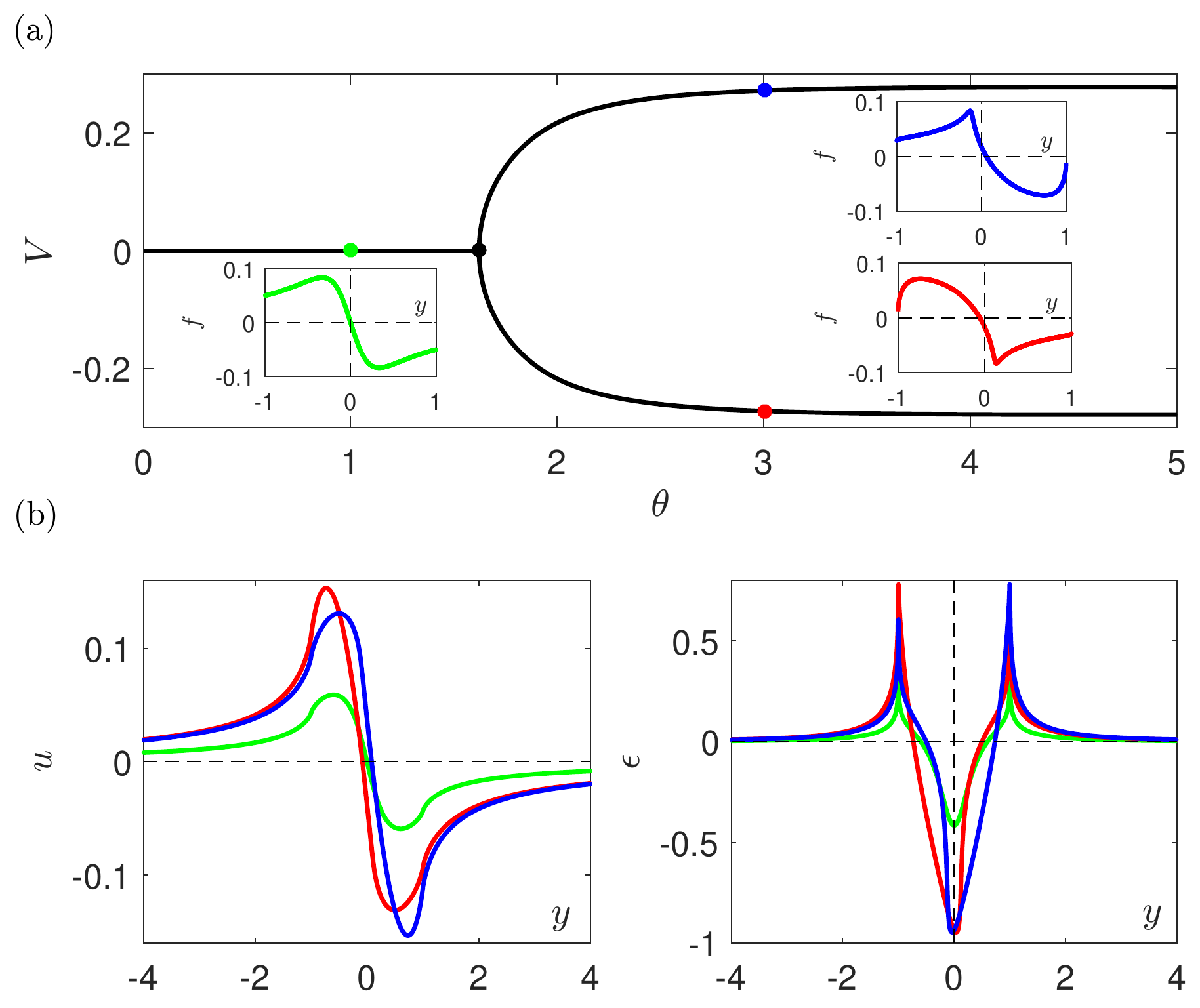}\hfill{}
\caption{ (a) Pitchfork bifurcation of the segment steady state velocity from a static to a motile state for the non-linear system \eqref{e:integ_pb}-\eqref{e:integ_cond}. Typical traction forces distributions for special choices of $\theta$ (same as for Fig.~\ref{f:pitchfork}) are shown in inset. (b) Substrate displacement and strain associated to the traction forces profiles shown in (a). Parameter $\beta=3$. }
\label{f:pitchfork_nl}
\end{figure} 

We compute by numerical continuation the traction force profile and the cell velocity beyond the bifurcation threshold for this more realistic sliding friction law, see Fig.~\ref{f:pitchfork_nl}. The traction force profiles are similar to the ones of the linear case except that their divergence at the trailing edge is eliminated because of the thresholding effect of $\Psi$. The cell-generated substrate displacement is also similar to the linear case, albeit smaller, leading to admissible strain profiles that satisfy $\epsilon>-1$.
\textcolor{black}{Using the model parameters estimated in Table~\ref{t:valpar}, we find at $\theta=2.5$ and $\beta=2$, a crawling velocity of the order of $\sim 0.1\mu\text{m~s}^{-1}$ similar to the one experimentally observed on stiff unconfined substrates ($\sim 0.2\mu\text{m~s}^{-1}$ \cite{rubinstein2009actin}). The typical order of magnitude of the traction forces that reach $\sim 50$ Pa is also comparable to the measurements performed in the aforementioned conditions which give $100$ Pa \cite{rubinstein2009actin}. While our model is not meant to reproduce quantitatively a specific experiment but rather to capture the physical ingredients necessary to observe an instability, matching the correct order of magnitude for these quantities suggests that the described spontaneous polarization mechanism is indeed applicable to cell motility.}

\section{Influence of large substrate deformations}\label{sec:large_def}

Equation \eqref{e:substrat_vel} and the relation $v_s=\partial_tu$ rely on the assumption that the  substrate deformation remains small. Even in the presence of a realistic non-linear friction law (finite $\beta$), the substrate strain can locally reach values where this simplifying assumption becomes questionable. See Fig.~\ref{f:pitchfork_nl}~(b). In this section, considering a different model for the elastic substrate amenable to a semi-analytic analysis, we show that the  protrusion-driven spontaneous motility initiation that we have evidenced in the previous sections occurs following an instability that is qualitatively unchanged when allowing large substrate deformations. 

To do so, instead of relating the substrate displacement to the applied traction forces using  formula \eqref{e:substrat_vel}, we consider a Winkler--Pasternak type foundation model \cite{pasternak1954new} where the ECM substrate is an elastic film subjected to tangential loading and elastically adhering to a stiff foundation, leading to
$$-E_s\Phi'(1+\partial_Xu)\partial_{XX}u+\frac{G}{l^2}u=f,$$  
where $E_s$ is the modulus of the film modeling the ECM, $\Phi$ is the non-linear stress-strain dependence ($'$ denotes its derivative), $l$ the film thickness and $G$ is the shear modulus of the connection to the foundation.  Importantly, $X$ is the Lagrangian spatial coordinate along the track in the reference (stress-free) configuration of the substrate. As a consequence, in Eulerian variables (current configuration) we have, $\partial_Xu=(1+\partial_Xu)\partial_xu$ and $v_s=\partial_tu+v_s\partial_xu$.
Thus, the strain being $\epsilon=\partial_Xu$ we have,
$$\epsilon=\frac{\partial_xu}{1-\partial_xu} \quad\text{and}\quad v_s=\frac{\partial_tu}{1-\partial_xu}.$$
In order to allow large deformations, we use the true strain $\Phi(1+\epsilon)=\log(1+\epsilon)$ as this choice naturally restricts deformations to the admissible domain, $\epsilon>-1$. We obtain,
\begin{equation}\label{e:stress_disp_film}
\left\lbrace 
\begin{array}{c}
-E_s\partial_x\epsilon+\frac{G}{l^2}u=f\\
(1+\epsilon)\partial_xu=\epsilon.
\end{array}
\right. 
\end{equation}
The above non-linear relation between $f$ and $u$ replaces \eqref{e:substrat_vel}. As a result, in non-dimensional form and with the change of variable from $(x,t)$ to $(y,s)$, the analogue of  the model of small deformation of a semi-infinite substrate  \eqref{e:integ_pb} is:
\begin{equation}\label{e:large_def_pb}
\left\lbrace 
\begin{array}{c}
-2\partial_y\epsilon+r\alpha^2u=\frac{\pi\alpha\theta}{6\beta}\Psi\left(2\beta(\dot{C}-y/2-(1+\epsilon)(\partial_su-2\dot{C}\partial_yu)) \right)\mathbf{1}_{[-1,1]}(y)\\
2(1+\epsilon)\partial_yu=\epsilon,
\end{array}
\right. 
\end{equation}
 where the additional non-dimensional parameters
 $$r=\frac{G}{E_s}\quad\text{and}\quad \alpha=\frac{L}{l}$$
represent the shear adhesion strength of the substrate to the foundation compared to the film elasticity and the cell length compared to the film thickness. The notation $\mathbf{1}_{[-1,1]}$ denotes the indicator function of the segment $[-1,1]$ where the traction forces are non vanishing. The substrate displacement $u$ is continuous across these points as well as the strain $\epsilon$ (due to stress continuity in the substrate). At $y=\pm \infty$, we impose the static conditions $u=0$ and $\epsilon=0$. Solving \eqref{e:large_def_pb} in quadrature in the two side domains where the traction forces vanish and using the continuity conditions, we reduce \eqref{e:large_def_pb} to the interval $[-1,1]$ with the appropriate boundary conditions %
and seek for traveling wave solutions for which $\dot{C}=V$ is a constant and $\partial_s u\equiv 0$:
\begin{equation}\label{e:large_def_pb_int}
\left\lbrace 
\begin{array}{c}
-2\partial_y\epsilon+r\alpha^2u=
  \frac{\pi\alpha\theta}{6\beta}
  \Psi\left(
     2\beta\left( (1+\epsilon)V-y/2\right) \right)\\
2(1+\epsilon)\partial_yu=\epsilon\\
r\alpha^2 u(\pm 1)^2/2=\epsilon(\pm 1)-\log (1+\epsilon(\pm 1)).
\end{array}
\right. 
\end{equation}
We solve \eqref{e:large_def_pb_int}-\eqref{e:integ_cond} for $(V,u,\epsilon)$
by numerical continuation \cite{doedel1981auto} starting from the trivial solution $V=0$, $u\equiv 0$ and $\epsilon\equiv 0$ when $\theta=0$. 
\begin{figure}[!t]
\hfill{}\includegraphics [scale=0.6]{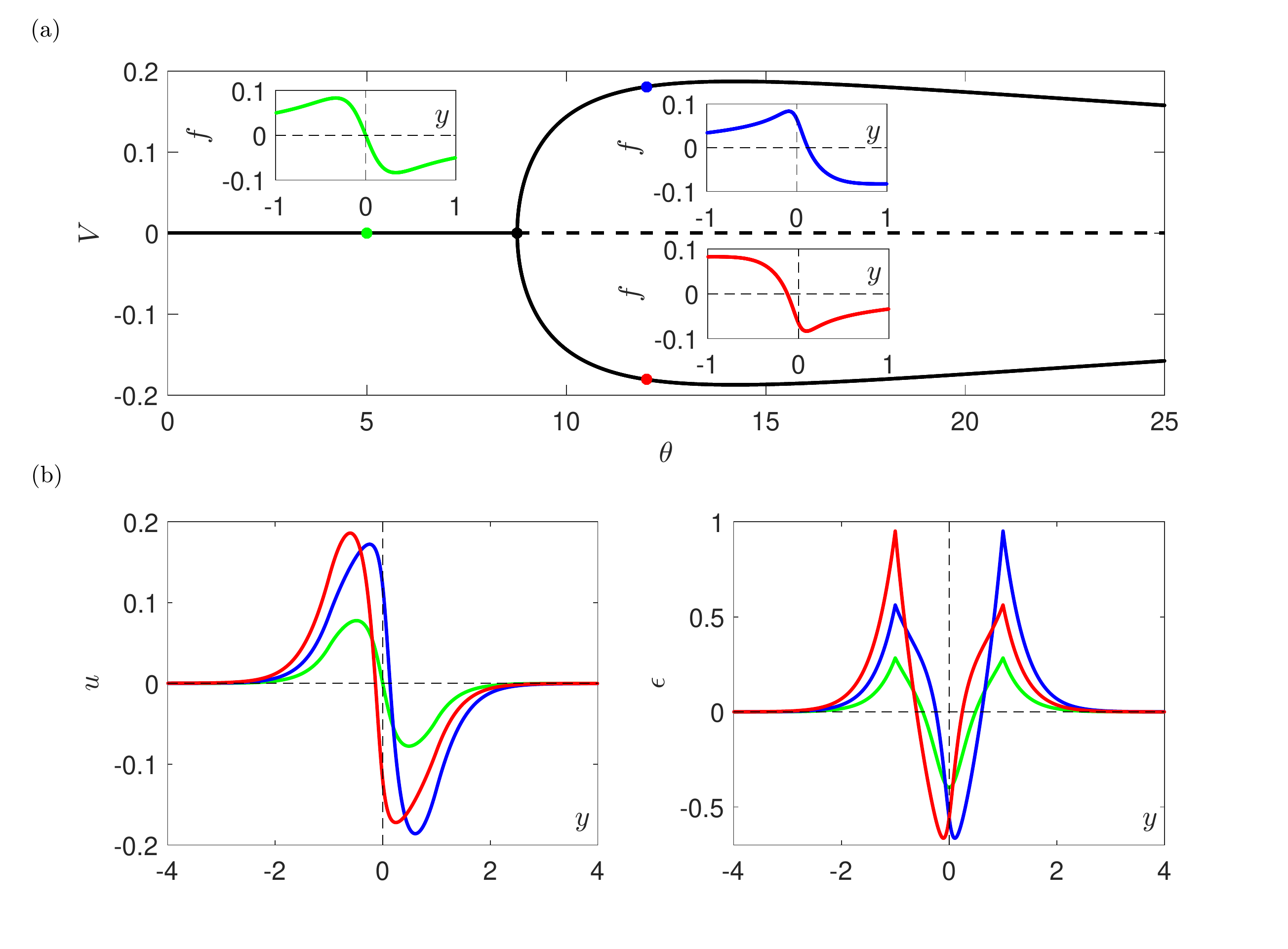}\hfill{}
\caption{(a) Bifurcation from a static to a motile state at a critical value of $\theta$ with the simplified model of the elastic foundation \eqref{e:large_def_pb_int}. We show in insets the traction forces profiles along the three branches at the points labeled with the related colored dots. (b) Substrate displacement and deformation along the three branches (color corresponds to the dots in (a)). Parameters $\beta=3$, $r=0.3$ and $\alpha=10$.}
\label{f:bif_diag_ld}
\end{figure} 

The situation is qualitatively similar to the one shown in Fig.~\ref{f:pitchfork_nl} when the substrate is modeled as an elastic incompressible semi-infinite material under small deformations: there exists a critical value of $\theta$ at which the symmetric and static state loses its stability to initiate an asymmetric traveling solution corresponding to a motile state.  The obtained traction forces and substrate displacement profiles are also comparable, showing that the cell polarization instability that we describe in this paper is consistently due to mechanisms independent of the small deformation assumption. Interestingly, in the post-bifurcated regime, the velocity dependence as a function of the substrate stiffness is not \textcolor{black}{monotonic} and displays a maximum for a particular substrate stiffness as observed in some experiments \cite{peyton2005extracellular}. Together with \cite{chelly2022cell}, these results indicate that this experimental observation may be explained by the emergent effect of non-local coupling of traction forces via the compliant substrate.

\begin{figure}[!t]
\hfill{}\includegraphics [scale=0.6]{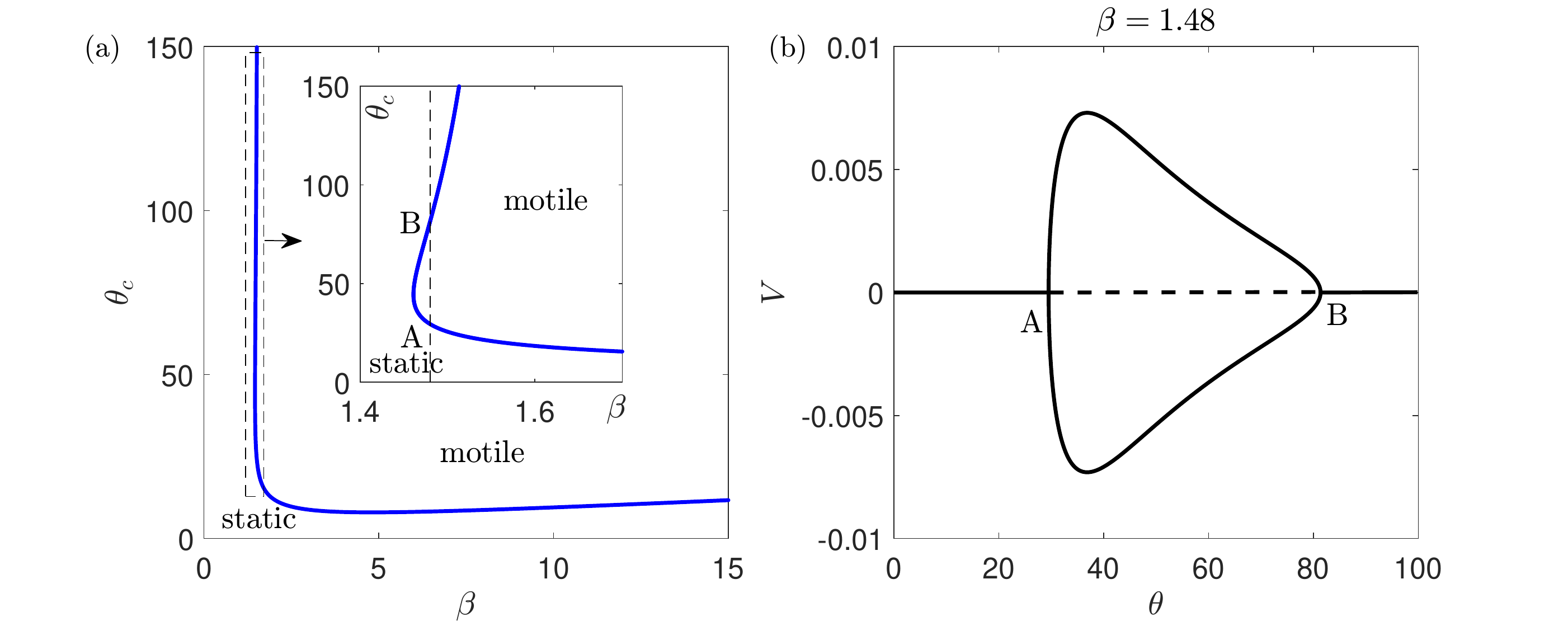}\hfill{}
\caption{(a) Locus of the bifurcation point between the static and motile solutions in the parameter space $(\beta,\theta)$. Inset shows the turning point. Points A and B are related to the bifurcation diagram shown on the right panel. (b) Bifurcation diagram showing an isola center type bifurcation at $\beta=1.48$ where the segment first polarizes at point A and then depolarizes at point B while the control parameter $\theta$ keeps increasing. Parameters $\alpha=10$ and $r=0.3$. }
\label{f:phase_diag_ld}
\end{figure}

Furthermore, we show in Fig.~\ref{f:phase_diag_ld}~(a) the dependence of the critical value $\theta_c$ at the bifurcation threshold for fixed parameters $\alpha$ and $r$.
As expected, when $\beta$ becomes sufficiently small, corresponding to a linear friction regime, we do not find any critical $\theta_c$ corresponding to the motility initiation instability, contrarily to what is observed in Sec.~\ref{sec:lin_fric}. This is because in the model \eqref{e:large_def_pb_int} rules out matter self-penetration, which according to condition~\eqref{e:surfing_condition} is necessary to obtain surfing with a linear friction. Instead, we observe a turning point in the dependence of $\theta_c$ on $\beta$ (Fig.~\ref{f:phase_diag_ld}~(a) inset) and the spontaneous motility instability is restricted to finite values of $\beta$. Due  to the existence of this turning point, for a range of $\beta$, the bifurcation can be of isola center type with a depolarization and reconnection to the static branch following the motility initiation. See Fig.~\ref{f:phase_diag_ld}~(b). The motility regime is therefore restricted to a finite range of substrate stiffness with neither too stiff nor too soft substrates enabling motility initiation in this case.  This qualitative result further underlines the potential role that non-linear \textcolor{black}{elasticity of the substrate} may play in the interpretation of experimental results showing a biphasic dependence of the cell velocity as a function of the substrate stiffness \cite{peyton2005extracellular}.

\section{Conclusions}

We have shown that symmetric protrusions driving the cell cytoskeleton flow can initiate spontaneous cell motility on a soft substrate. This instability 
\textcolor{black}{requires two well-established non-linearities that we have deliberately selected among others to obtain our minimal model. 
First, a geometric non-linearity that finds its origin in the advection of the substrate strain with the cell motion. Thereby, the cell surfs on the substrate deformation it creates. The feedback between the substrate and the cell mechanics is fully mediated by the frictional traction forces and requires no specific mechanosensing, which suggests that this phenomenon could be universal across cell types. However, the physical admissibility of the motile configuration requires 
a second non-linearity of material type in the effective sliding friction interaction between the cell and the substrate.
 This experimentally established biphasic behavior \cite{gardel2008traction} eliminates the spurious matter self-penetration beyond the instability threshold, although this ingredient is not needed to understand the instability in itself.} Considering large deformations in the substrate confirms the robustness of the mechanism of the cell polarization transition and unravels a more complex bifurcation diagram than in the small deformation case, with the possibility of a biphasic dependence of the cell velocity as a function of the substrate stiffness.  

Even though experimentally challenging, future work could aim at deciding whether the substrate compliance increases significantly the frequency of motility initiation as it was done in \cite{barnhart2015balance} to investigate the role of various drugs on motility initiation. Indeed, on glass, constraining keratocytes to move along narrow adhesive  $5 \mu$m tracks led to circa 98\% of immobile cells  in  \cite{Mohammed+Gabriele.2019.1}. Combining soft substrates and narrow tracks could thus allow to test our model. \textcolor{black}{Qualitatively, the bifurcation threshold \eqref{e:theta_c_frict_nl} indeed states that when the magnitude of the traction forces due to protrusive stress becomes of the same order as the substrate stiffness, there is a transition from a static to a motile state. As Traction Force Miscroscopy allows to quantify these stresses \cite{Mohammed+Gabriele.2019.1} and microfabrication makes it possible to produce substrates with a 
corresponding elastic modulus \cite{chelly2022cell}, this suggests a path to experimentally find a critical substrate softness beyond which a given cell type starts to move along a track.}

\textcolor{black}{To be more quantitative and comprehensive, this model could be enriched with other mechanisms of self-polarization on one dimensional tracks, such as contraction-driven motility \cite{recho2013contraction} that can also operate for infinitely stiff substrates. It could also be coupled to internal cell signaling, controlling the local protrusion rates at the cell boundary as a function of  Rho GTPases concentration that undergo reaction-drift-diffusion dynamics in the cytoplasm \cite{mori2008wave}, further enhancing the cell polarity when the symmetry of the internal flow is broken \cite{Ambrosi-Zanzottera.2016.1,Ron+Gov.2020.1}.} 

Another interesting generalization to reach a more quantitative model of a specific experiment would be to represent the substrate rheology and the track geometry in a refined way \cite{chelly2022cell}. \textcolor{black}{Finally, the coupling between fluctuations of the internal cytoskeleton dynamics and the confining geometry \cite{buskermolen2019entropic} and stiffness \cite{ippolito2021contact} of the cell environment is also known to play an important role in cell motility.}

\section*{Acknowledgments}
The authors thank L. Truskinovsky for his stimulating comments, as well as Alexander Erlich for his critical and helpful reading of the manuscript.


\bibliographystyle{elsarticle-harv}
\addcontentsline{toc}{section}{\refname}

\bibliography{substrate_turnover_polarity,jocelyn}

\begin{thebibliography}{66}
\expandafter\ifx\csname natexlab\endcsname\relax\def\natexlab#1{#1}\fi
\providecommand{\url}[1]{\texttt{#1}}
\providecommand{\href}[2]{#2}
\providecommand{\path}[1]{#1}
\providecommand{\DOIprefix}{doi:}
\providecommand{\ArXivprefix}{arXiv:}
\providecommand{\URLprefix}{URL: }
\providecommand{\Pubmedprefix}{pmid:}
\providecommand{\doi}[1]{\href{http://dx.doi.org/#1}{\path{#1}}}
\providecommand{\Pubmed}[1]{\href{pmid:#1}{\path{#1}}}
\providecommand{\bibinfo}[2]{#2}
\ifx\xfnm\relax \def\xfnm[#1]{\unskip,\space#1}\fi
\bibitem[{Ambrosi and Zanzottera(2016)}]{Ambrosi-Zanzottera.2016.1}
\bibinfo{author}{Ambrosi, D.}, \bibinfo{author}{Zanzottera, A.},
  \bibinfo{year}{2016}.
\newblock \bibinfo{title}{Mechanics and polarity in cell motility}.
\newblock \bibinfo{journal}{Physica D} ,
  \bibinfo{pages}{58--66}\DOIprefix\doi{10.1016/j.physd.2016.05.003}.
\bibitem[{Banerjee and Marchetti(2011)}]{banerjee2011substrate}
\bibinfo{author}{Banerjee, S.}, \bibinfo{author}{Marchetti, M.C.},
  \bibinfo{year}{2011}.
\newblock \bibinfo{title}{Substrate rigidity deforms and polarizes active
  gels}.
\newblock \bibinfo{journal}{EPL (Europhysics Letters)} \bibinfo{volume}{96},
  \bibinfo{pages}{28003}.
\bibitem[{Barnhart et~al.(2015)Barnhart, Lee, Allen, Theriot and
  Mogilner}]{barnhart2015balance}
\bibinfo{author}{Barnhart, E.}, \bibinfo{author}{Lee, K.C.},
  \bibinfo{author}{Allen, G.M.}, \bibinfo{author}{Theriot, J.A.},
  \bibinfo{author}{Mogilner, A.}, \bibinfo{year}{2015}.
\newblock \bibinfo{title}{Balance between cell- substrate adhesion and myosin
  contraction determines the frequency of motility initiation in fish
  keratocytes}.
\newblock \bibinfo{journal}{Proceedings of the National Academy of Sciences}
  \bibinfo{volume}{112}, \bibinfo{pages}{5045--5050}.
\bibitem[{Barnhart et~al.(2011)Barnhart, Lee, Keren, Mogilner and
  Theriot}]{barnhart2011adhesion}
\bibinfo{author}{Barnhart, E.L.}, \bibinfo{author}{Lee, K.C.},
  \bibinfo{author}{Keren, K.}, \bibinfo{author}{Mogilner, A.},
  \bibinfo{author}{Theriot, J.A.}, \bibinfo{year}{2011}.
\newblock \bibinfo{title}{An adhesion-dependent switch between mechanisms that
  determine motile cell shape}.
\newblock \bibinfo{journal}{PLoS biology} \bibinfo{volume}{9},
  \bibinfo{pages}{e1001059}.
\bibitem[{Bergert et~al.(2015)Bergert, Erzberger, Desai, Aspalter, Oates,
  Charras, Salbreux and Paluch}]{bergert2015force}
\bibinfo{author}{Bergert, M.}, \bibinfo{author}{Erzberger, A.},
  \bibinfo{author}{Desai, R.A.}, \bibinfo{author}{Aspalter, I.M.},
  \bibinfo{author}{Oates, A.C.}, \bibinfo{author}{Charras, G.},
  \bibinfo{author}{Salbreux, G.}, \bibinfo{author}{Paluch, E.K.},
  \bibinfo{year}{2015}.
\newblock \bibinfo{title}{Force transmission during adhesion-independent
  migration}.
\newblock \bibinfo{journal}{Nature cell biology} \bibinfo{volume}{17},
  \bibinfo{pages}{524--529}.
\bibitem[{Bigoni and Noselli(2011)}]{bigoni2011experimental}
\bibinfo{author}{Bigoni, D.}, \bibinfo{author}{Noselli, G.},
  \bibinfo{year}{2011}.
\newblock \bibinfo{title}{Experimental evidence of flutter and divergence
  instabilities induced by dry friction}.
\newblock \bibinfo{journal}{Journal of the Mechanics and Physics of Solids}
  \bibinfo{volume}{59}, \bibinfo{pages}{2208--2226}.
\bibitem[{Blanch-Mercader and Casademunt(2013)}]{blanch2013spontaneous}
\bibinfo{author}{Blanch-Mercader, C.}, \bibinfo{author}{Casademunt, J.},
  \bibinfo{year}{2013}.
\newblock \bibinfo{title}{Spontaneous motility of actin lamellar fragments}.
\newblock \bibinfo{journal}{Physical review letters} \bibinfo{volume}{110},
  \bibinfo{pages}{078102}.
\bibitem[{Boal(2012)}]{boal_2012}
\bibinfo{author}{Boal, D.}, \bibinfo{year}{2012}.
\newblock \bibinfo{title}{Mechanics of the Cell}.
\newblock \bibinfo{edition}{2} ed., \bibinfo{publisher}{Cambridge University
  Press}.
\newblock \DOIprefix\doi{10.1017/CBO9781139022217}.
\bibitem[{Buskermolen et~al.(2019)Buskermolen, Suresh, Shishvan, Vigliotti,
  DeSimone, Kurniawan, Bouten and Deshpande}]{buskermolen2019entropic}
\bibinfo{author}{Buskermolen, A.B.}, \bibinfo{author}{Suresh, H.},
  \bibinfo{author}{Shishvan, S.S.}, \bibinfo{author}{Vigliotti, A.},
  \bibinfo{author}{DeSimone, A.}, \bibinfo{author}{Kurniawan, N.A.},
  \bibinfo{author}{Bouten, C.V.}, \bibinfo{author}{Deshpande, V.S.},
  \bibinfo{year}{2019}.
\newblock \bibinfo{title}{Entropic forces drive cellular contact guidance}.
\newblock \bibinfo{journal}{Biophysical journal} \bibinfo{volume}{116},
  \bibinfo{pages}{1994--2008}.
\bibitem[{Callan-Jones and Voituriez(2013)}]{callan2013active}
\bibinfo{author}{Callan-Jones, A.}, \bibinfo{author}{Voituriez, R.},
  \bibinfo{year}{2013}.
\newblock \bibinfo{title}{Active gel model of amoeboid cell motility}.
\newblock \bibinfo{journal}{New Journal of Physics} \bibinfo{volume}{15},
  \bibinfo{pages}{025022}.
\bibitem[{Carlsson(2011)}]{carlsson2011mechanisms}
\bibinfo{author}{Carlsson, A.}, \bibinfo{year}{2011}.
\newblock \bibinfo{title}{Mechanisms of cell propulsion by active stresses}.
\newblock \bibinfo{journal}{New journal of physics} \bibinfo{volume}{13},
  \bibinfo{pages}{073009}.
\bibitem[{Chelly et~al.(2022)Chelly, Jahangiri, Mireux, {\'E}tienne, Dysthe,
  Verdier and Recho}]{chelly2022cell}
\bibinfo{author}{Chelly, H.}, \bibinfo{author}{Jahangiri, A.},
  \bibinfo{author}{Mireux, M.}, \bibinfo{author}{{\'E}tienne, J.},
  \bibinfo{author}{Dysthe, D.}, \bibinfo{author}{Verdier, C.},
  \bibinfo{author}{Recho, P.}, \bibinfo{year}{2022}.
\newblock \bibinfo{title}{Cell crawling on a compliant substrate: a biphasic
  relation with linear friction}.
\newblock \bibinfo{journal}{International Journal of Non-Linear Mechanics}
  \bibinfo{volume}{139}, \bibinfo{pages}{103897}.
\bibitem[{Chen et~al.(2022)Chen, Feng and Li}]{chen2022unified}
\bibinfo{author}{Chen, P.C.}, \bibinfo{author}{Feng, X.Q.},
  \bibinfo{author}{Li, B.}, \bibinfo{year}{2022}.
\newblock \bibinfo{title}{Unified multiscale theory of cellular mechanical
  adaptations to substrate stiffness}.
\newblock \bibinfo{journal}{Biophysical Journal} \bibinfo{volume}{121},
  \bibinfo{pages}{3474--3485}.
\bibitem[{Cicconofri and DeSimone(2015)}]{cicconofri2015motility}
\bibinfo{author}{Cicconofri, G.}, \bibinfo{author}{DeSimone, A.},
  \bibinfo{year}{2015}.
\newblock \bibinfo{title}{Motility of a model bristle-bot: A theoretical
  analysis}.
\newblock \bibinfo{journal}{International Journal of Non-Linear Mechanics}
  \bibinfo{volume}{76}, \bibinfo{pages}{233--239}.
\bibitem[{Copos and Mogilner(2020)}]{copos2020hybrid}
\bibinfo{author}{Copos, C.}, \bibinfo{author}{Mogilner, A.},
  \bibinfo{year}{2020}.
\newblock \bibinfo{title}{A hybrid stochastic--deterministic mechanochemical
  model of cell polarization}.
\newblock \bibinfo{journal}{Molecular biology of the cell}
  \bibinfo{volume}{31}, \bibinfo{pages}{1637--1649}.
\bibitem[{DeSimone and Tatone(2012)}]{desimone2012crawling}
\bibinfo{author}{DeSimone, A.}, \bibinfo{author}{Tatone, A.},
  \bibinfo{year}{2012}.
\newblock \bibinfo{title}{Crawling motility through the analysis of model
  locomotors: two case studies}.
\newblock \bibinfo{journal}{The European Physical Journal E}
  \bibinfo{volume}{35}, \bibinfo{pages}{1--8}.
\bibitem[{Doedel(1981)}]{doedel1981auto}
\bibinfo{author}{Doedel, E.J.}, \bibinfo{year}{1981}.
\newblock \bibinfo{title}{Auto: A program for the automatic bifurcation
  analysis of autonomous systems}.
\newblock \bibinfo{journal}{Congr. Numer} \bibinfo{volume}{30},
  \bibinfo{pages}{25--93}.
\bibitem[{Dolega et~al.(2021)Dolega, Zurlo, Le~Goff, Greda, Verdier, Joanny,
  Cappello and Recho}]{dolega2021mechanical}
\bibinfo{author}{Dolega, M.}, \bibinfo{author}{Zurlo, G.},
  \bibinfo{author}{Le~Goff, M.}, \bibinfo{author}{Greda, M.},
  \bibinfo{author}{Verdier, C.}, \bibinfo{author}{Joanny, J.F.},
  \bibinfo{author}{Cappello, G.}, \bibinfo{author}{Recho, P.},
  \bibinfo{year}{2021}.
\newblock \bibinfo{title}{Mechanical behavior of multi-cellular spheroids under
  osmotic compression}.
\newblock \bibinfo{journal}{Journal of the Mechanics and Physics of Solids}
  \bibinfo{volume}{147}, \bibinfo{pages}{104205}.
\bibitem[{Doyle et~al.(2013)Doyle, Petrie, Kutys and
  Yamada}]{doyle2013dimensions}
\bibinfo{author}{Doyle, A.D.}, \bibinfo{author}{Petrie, R.J.},
  \bibinfo{author}{Kutys, M.L.}, \bibinfo{author}{Yamada, K.M.},
  \bibinfo{year}{2013}.
\newblock \bibinfo{title}{Dimensions in cell migration}.
\newblock \bibinfo{journal}{Current opinion in cell biology}
  \bibinfo{volume}{25}, \bibinfo{pages}{642--649}.
\bibitem[{Doyle et~al.(2009)Doyle, Wang, Matsumoto and Yamada}]{doyle2009one}
\bibinfo{author}{Doyle, A.D.}, \bibinfo{author}{Wang, F.W.},
  \bibinfo{author}{Matsumoto, K.}, \bibinfo{author}{Yamada, K.M.},
  \bibinfo{year}{2009}.
\newblock \bibinfo{title}{One-dimensional topography underlies
  three-dimensional fibrillar cell migration}.
\newblock \bibinfo{journal}{The Journal of cell biology} \bibinfo{volume}{184},
  \bibinfo{pages}{481--490}.
\bibitem[{Drozdowski et~al.(2021)Drozdowski, Ziebert and
  Schwarz}]{Drozdowski+Schwarz.2021.1}
\bibinfo{author}{Drozdowski, O.M.}, \bibinfo{author}{Ziebert, F.},
  \bibinfo{author}{Schwarz, U.S.}, \bibinfo{year}{2021}.
\newblock \bibinfo{title}{Optogenetic control of intracellular flows and cell
  migration: A comprehensive mathematical analysis with a minimal active gel
  model}.
\newblock \bibinfo{journal}{Phys. Rev. E} \bibinfo{volume}{104}.
\newblock \DOIprefix\doi{10.1103/PhysRevE.104.024406}.
\bibitem[{DuChez et~al.(2019)DuChez, Doyle, Dimitriadis and
  Yamada}]{duchez2019durotaxis}
\bibinfo{author}{DuChez, B.J.}, \bibinfo{author}{Doyle, A.D.},
  \bibinfo{author}{Dimitriadis, E.K.}, \bibinfo{author}{Yamada, K.M.},
  \bibinfo{year}{2019}.
\newblock \bibinfo{title}{Durotaxis by human cancer cells}.
\newblock \bibinfo{journal}{Biophysical journal} \bibinfo{volume}{116},
  \bibinfo{pages}{670--683}.
\bibitem[{Estrada and Kanwal(1989)}]{estrada1989integral}
\bibinfo{author}{Estrada, R.}, \bibinfo{author}{Kanwal, R.},
  \bibinfo{year}{1989}.
\newblock \bibinfo{title}{Integral equations with logarithmic kernels}.
\newblock \bibinfo{journal}{IMA journal of applied mathematics}
  \bibinfo{volume}{43}, \bibinfo{pages}{133--155}.
\bibitem[{Feng et~al.(2019)Feng, Levine, Mao and Sander}]{feng2019cell}
\bibinfo{author}{Feng, J.}, \bibinfo{author}{Levine, H.}, \bibinfo{author}{Mao,
  X.}, \bibinfo{author}{Sander, L.M.}, \bibinfo{year}{2019}.
\newblock \bibinfo{title}{Cell motility, contact guidance, and durotaxis}.
\newblock \bibinfo{journal}{Soft matter} \bibinfo{volume}{15},
  \bibinfo{pages}{4856--4864}.
\bibitem[{Gardel et~al.(2008)Gardel, Sabass, Ji, Danuser, Schwarz and
  Waterman}]{gardel2008traction}
\bibinfo{author}{Gardel, M.L.}, \bibinfo{author}{Sabass, B.},
  \bibinfo{author}{Ji, L.}, \bibinfo{author}{Danuser, G.},
  \bibinfo{author}{Schwarz, U.S.}, \bibinfo{author}{Waterman, C.M.},
  \bibinfo{year}{2008}.
\newblock \bibinfo{title}{Traction stress in focal adhesions correlates
  biphasically with actin retrograde flow speed}.
\newblock \bibinfo{journal}{The Journal of cell biology} \bibinfo{volume}{183},
  \bibinfo{pages}{999--1005}.
\bibitem[{Giomi and DeSimone(2014)}]{giomi2014spontaneous}
\bibinfo{author}{Giomi, L.}, \bibinfo{author}{DeSimone, A.},
  \bibinfo{year}{2014}.
\newblock \bibinfo{title}{Spontaneous division and motility in active nematic
  droplets}.
\newblock \bibinfo{journal}{Physical review letters} \bibinfo{volume}{112},
  \bibinfo{pages}{147802}.
\bibitem[{Hawkins et~al.(2009)Hawkins, Piel, Faure-Andre, Lennon-Dumenil,
  Joanny, Prost and Voituriez}]{hawkins2009pushing}
\bibinfo{author}{Hawkins, R.J.}, \bibinfo{author}{Piel, M.},
  \bibinfo{author}{Faure-Andre, G.}, \bibinfo{author}{Lennon-Dumenil, A.},
  \bibinfo{author}{Joanny, J.}, \bibinfo{author}{Prost, J.},
  \bibinfo{author}{Voituriez, R.}, \bibinfo{year}{2009}.
\newblock \bibinfo{title}{Pushing off the walls: a mechanism of cell motility
  in confinement}.
\newblock \bibinfo{journal}{Physical review letters} \bibinfo{volume}{102},
  \bibinfo{pages}{058103}.
\bibitem[{Hennig et~al.(2020)Hennig, Wang, Moreau, Valon, DeBeco, Coppey,
  Miroshnikova, Albiges-Rizo, Favard, Voituriez et~al.}]{hennig2020stick}
\bibinfo{author}{Hennig, K.}, \bibinfo{author}{Wang, I.},
  \bibinfo{author}{Moreau, P.}, \bibinfo{author}{Valon, L.},
  \bibinfo{author}{DeBeco, S.}, \bibinfo{author}{Coppey, M.},
  \bibinfo{author}{Miroshnikova, Y.}, \bibinfo{author}{Albiges-Rizo, C.},
  \bibinfo{author}{Favard, C.}, \bibinfo{author}{Voituriez, R.}, et~al.,
  \bibinfo{year}{2020}.
\newblock \bibinfo{title}{Stick-slip dynamics of cell adhesion triggers
  spontaneous symmetry breaking and directional migration of mesenchymal cells
  on one-dimensional lines}.
\newblock \bibinfo{journal}{Science advances} \bibinfo{volume}{6},
  \bibinfo{pages}{eaau5670}.
\bibitem[{Ippolito and Deshpande(2021)}]{ippolito2021contact}
\bibinfo{author}{Ippolito, A.}, \bibinfo{author}{Deshpande, V.S.},
  \bibinfo{year}{2021}.
\newblock \bibinfo{title}{Contact guidance via heterogeneity of substrate
  elasticity}.
\newblock \bibinfo{journal}{Acta Biomaterialia} .
\bibitem[{Johnson(1987)}]{Johnson1987}
\bibinfo{author}{Johnson, K.L.}, \bibinfo{year}{1987}.
\newblock \bibinfo{title}{Contact mechanics}.
\newblock \bibinfo{publisher}{Cambridge university press}.
\bibitem[{Julicher et~al.(2007)Julicher, Kruse, Prost and
  Joanny}]{Julicher+Joanny.2007.1}
\bibinfo{author}{Julicher, F.}, \bibinfo{author}{Kruse, K.},
  \bibinfo{author}{Prost, J.}, \bibinfo{author}{Joanny, J.},
  \bibinfo{year}{2007}.
\newblock \bibinfo{title}{Active behavior of the cytoskeleton}.
\newblock \bibinfo{journal}{Physics Reports} \bibinfo{volume}{449},
  \bibinfo{pages}{3--28}.
\newblock \DOIprefix\doi{10.1016/j.physrep.2007.02.018}.
\bibitem[{Karpenko(1966)}]{Karpenko1966approximate}
\bibinfo{author}{Karpenko, L.}, \bibinfo{year}{1966}.
\newblock \bibinfo{title}{Approximate solution of a singular integral equation
  by means of jacobi polynomials}.
\newblock \bibinfo{journal}{Journal of Applied Mathematics and Mechanics}
  \bibinfo{volume}{30}, \bibinfo{pages}{668--675}.
\newblock \URLprefix
  \url{https://www.sciencedirect.com/science/article/pii/0021892867901037},
  \DOIprefix\doi{https://doi.org/10.1016/0021-8928(67)90103-7}.
\bibitem[{Kruse et~al.(2006)Kruse, Joanny, J{\"u}licher and
  Prost}]{kruse2006contractility}
\bibinfo{author}{Kruse, K.}, \bibinfo{author}{Joanny, J.},
  \bibinfo{author}{J{\"u}licher, F.}, \bibinfo{author}{Prost, J.},
  \bibinfo{year}{2006}.
\newblock \bibinfo{title}{Contractility and retrograde flow in lamellipodium
  motion}.
\newblock \bibinfo{journal}{Physical biology} \bibinfo{volume}{3},
  \bibinfo{pages}{130}.
\bibitem[{Kruse et~al.(2005)Kruse, Joanny, J{\"u}licher, Prost and
  Sekimoto}]{kruse2005generic}
\bibinfo{author}{Kruse, K.}, \bibinfo{author}{Joanny, J.F.},
  \bibinfo{author}{J{\"u}licher, F.}, \bibinfo{author}{Prost, J.},
  \bibinfo{author}{Sekimoto, K.}, \bibinfo{year}{2005}.
\newblock \bibinfo{title}{Generic theory of active polar gels: a paradigm for
  cytoskeletal dynamics}.
\newblock \bibinfo{journal}{The European Physical Journal E}
  \bibinfo{volume}{16}, \bibinfo{pages}{5--16}.
\bibitem[{Landau and Lifshitz(1984)}]{Landau+.1984.8}
\bibinfo{author}{Landau, L.D.}, \bibinfo{author}{Lifshitz, E.},
  \bibinfo{year}{1984}.
\newblock \bibinfo{title}{Electrodynamics of Continuous Media}.
  volume~\bibinfo{volume}{8} of \textit{\bibinfo{series}{Course of Theoretical
  Physics}}.
\newblock \bibinfo{publisher}{Pergamon Press}, \bibinfo{address}{Oxford}.
\newblock \bibinfo{note}{2nd ed}.
\bibitem[{Larripa and Mogilner(2006)}]{larripa2006transport}
\bibinfo{author}{Larripa, K.}, \bibinfo{author}{Mogilner, A.},
  \bibinfo{year}{2006}.
\newblock \bibinfo{title}{Transport of a 1d viscoelastic actin--myosin strip of
  gel as a model of a crawling cell}.
\newblock \bibinfo{journal}{Physica A: Statistical Mechanics and its
  Applications} \bibinfo{volume}{372}, \bibinfo{pages}{113--123}.
\bibitem[{Laurent et~al.(2005)Laurent, Kasas, Yersin, Sch{\"a}ffer, Catsicas,
  Dietler, Verkhovsky and Meister}]{laurent2005gradient}
\bibinfo{author}{Laurent, V.M.}, \bibinfo{author}{Kasas, S.},
  \bibinfo{author}{Yersin, A.}, \bibinfo{author}{Sch{\"a}ffer, T.E.},
  \bibinfo{author}{Catsicas, S.}, \bibinfo{author}{Dietler, G.},
  \bibinfo{author}{Verkhovsky, A.B.}, \bibinfo{author}{Meister, J.J.},
  \bibinfo{year}{2005}.
\newblock \bibinfo{title}{Gradient of rigidity in the lamellipodia of migrating
  cells revealed by atomic force microscopy}.
\newblock \bibinfo{journal}{Biophysical journal} \bibinfo{volume}{89},
  \bibinfo{pages}{667--675}.
\bibitem[{Lavi et~al.(2020)Lavi, Meunier, Voituriez and
  Casademunt}]{lavi2020motility}
\bibinfo{author}{Lavi, I.}, \bibinfo{author}{Meunier, N.},
  \bibinfo{author}{Voituriez, R.}, \bibinfo{author}{Casademunt, J.},
  \bibinfo{year}{2020}.
\newblock \bibinfo{title}{Motility and morphodynamics of confined cells}.
\newblock \bibinfo{journal}{Physical Review E} \bibinfo{volume}{101},
  \bibinfo{pages}{022404}.
\bibitem[{Lelidis and Joanny(2013)}]{lelidis2013interaction}
\bibinfo{author}{Lelidis, I.}, \bibinfo{author}{Joanny, J.F.},
  \bibinfo{year}{2013}.
\newblock \bibinfo{title}{Interaction of focal adhesions mediated by the
  substrate elasticity}.
\newblock \bibinfo{journal}{Soft Matter} \bibinfo{volume}{9},
  \bibinfo{pages}{11120--11128}.
\bibitem[{Levental et~al.(2009)Levental, Yu, Kass, Lakins, Egeblad, Erler,
  Fong, Csiszar, Giaccia, Weninger et~al.}]{levental2009matrix}
\bibinfo{author}{Levental, K.R.}, \bibinfo{author}{Yu, H.},
  \bibinfo{author}{Kass, L.}, \bibinfo{author}{Lakins, J.N.},
  \bibinfo{author}{Egeblad, M.}, \bibinfo{author}{Erler, J.T.},
  \bibinfo{author}{Fong, S.F.}, \bibinfo{author}{Csiszar, K.},
  \bibinfo{author}{Giaccia, A.}, \bibinfo{author}{Weninger, W.}, et~al.,
  \bibinfo{year}{2009}.
\newblock \bibinfo{title}{Matrix crosslinking forces tumor progression by
  enhancing integrin signaling}.
\newblock \bibinfo{journal}{Cell} \bibinfo{volume}{139},
  \bibinfo{pages}{891--906}.
\bibitem[{Lin(2010)}]{Lin2010model}
\bibinfo{author}{Lin, Y.}, \bibinfo{year}{2010}.
\newblock \bibinfo{title}{A model of cell motility leading to biphasic
  dependence of transport speed on adhesive strength}.
\newblock \bibinfo{journal}{Journal of the Mechanics and Physics of Solids}
  \bibinfo{volume}{58}, \bibinfo{pages}{502--514}.
\bibitem[{Lo et~al.(2000)Lo, Wang, Dembo and Wang}]{lo2000cell}
\bibinfo{author}{Lo, C.M.}, \bibinfo{author}{Wang, H.B.},
  \bibinfo{author}{Dembo, M.}, \bibinfo{author}{Wang, Y.l.},
  \bibinfo{year}{2000}.
\newblock \bibinfo{title}{Cell movement is guided by the rigidity of the
  substrate}.
\newblock \bibinfo{journal}{Biophysical journal} \bibinfo{volume}{79},
  \bibinfo{pages}{144--152}.
\bibitem[{L{\"o}ber et~al.(2014)L{\"o}ber, Ziebert and
  Aranson}]{lober2014modeling}
\bibinfo{author}{L{\"o}ber, J.}, \bibinfo{author}{Ziebert, F.},
  \bibinfo{author}{Aranson, I.S.}, \bibinfo{year}{2014}.
\newblock \bibinfo{title}{Modeling crawling cell movement on soft engineered
  substrates}.
\newblock \bibinfo{journal}{Soft matter} \bibinfo{volume}{10},
  \bibinfo{pages}{1365--1373}.
\bibitem[{Maiuri et~al.(2012)Maiuri, Terriac, Paul-Gilloteaux, Vignaud,
  McNally, Onuffer, Thorn, Nguyen, Georgoulia, Soong et~al.}]{maiuri2012first}
\bibinfo{author}{Maiuri, P.}, \bibinfo{author}{Terriac, E.},
  \bibinfo{author}{Paul-Gilloteaux, P.}, \bibinfo{author}{Vignaud, T.},
  \bibinfo{author}{McNally, K.}, \bibinfo{author}{Onuffer, J.},
  \bibinfo{author}{Thorn, K.}, \bibinfo{author}{Nguyen, P.A.},
  \bibinfo{author}{Georgoulia, N.}, \bibinfo{author}{Soong, D.}, et~al.,
  \bibinfo{year}{2012}.
\newblock \bibinfo{title}{The first world cell race}.
\newblock \bibinfo{journal}{Current Biology} \bibinfo{volume}{22},
  \bibinfo{pages}{R673--R675}.
\bibitem[{Mohammed et~al.(2019)Mohammed, Charras, Vercruysse, Versaevel,
  Lantoine, Alaimo, Bruy{\`e}re, Luciano, Glinel, Delhaye
  et~al.}]{Mohammed+Gabriele.2019.1}
\bibinfo{author}{Mohammed, D.}, \bibinfo{author}{Charras, G.},
  \bibinfo{author}{Vercruysse, E.}, \bibinfo{author}{Versaevel, M.},
  \bibinfo{author}{Lantoine, J.}, \bibinfo{author}{Alaimo, L.},
  \bibinfo{author}{Bruy{\`e}re, C.}, \bibinfo{author}{Luciano, M.},
  \bibinfo{author}{Glinel, K.}, \bibinfo{author}{Delhaye, G.}, et~al.,
  \bibinfo{year}{2019}.
\newblock \bibinfo{title}{Substrate area confinement is a key determinant of
  cell velocity in collective migration}.
\newblock \bibinfo{journal}{Nature Physics} \bibinfo{volume}{15},
  \bibinfo{pages}{858--866}.
\bibitem[{Mori et~al.(2008)Mori, Jilkine and Edelstein-Keshet}]{mori2008wave}
\bibinfo{author}{Mori, Y.}, \bibinfo{author}{Jilkine, A.},
  \bibinfo{author}{Edelstein-Keshet, L.}, \bibinfo{year}{2008}.
\newblock \bibinfo{title}{Wave-pinning and cell polarity from a bistable
  reaction-diffusion system}.
\newblock \bibinfo{journal}{Biophysical journal} \bibinfo{volume}{94},
  \bibinfo{pages}{3684--3697}.
\bibitem[{Oliveri et~al.(2021)Oliveri, Franze and Goriely}]{oliveri2021theory}
\bibinfo{author}{Oliveri, H.}, \bibinfo{author}{Franze, K.},
  \bibinfo{author}{Goriely, A.}, \bibinfo{year}{2021}.
\newblock \bibinfo{title}{Theory for durotactic axon guidance}.
\newblock \bibinfo{journal}{Physical Review Letters} \bibinfo{volume}{126},
  \bibinfo{pages}{118101}.
\bibitem[{Pasternak(1954)}]{pasternak1954new}
\bibinfo{author}{Pasternak, P.}, \bibinfo{year}{1954}.
\newblock \bibinfo{title}{On a new method of analysis of an elastic foundation
  by means of two foundation constants}.
\newblock \bibinfo{journal}{Gos. Izd. Lit. po Strait i Arkh} .
\bibitem[{Peyton and Putnam(2005)}]{peyton2005extracellular}
\bibinfo{author}{Peyton, S.R.}, \bibinfo{author}{Putnam, A.J.},
  \bibinfo{year}{2005}.
\newblock \bibinfo{title}{Extracellular matrix rigidity governs smooth muscle
  cell motility in a biphasic fashion}.
\newblock \bibinfo{journal}{Journal of cellular physiology}
  \bibinfo{volume}{204}, \bibinfo{pages}{198--209}.
\bibitem[{Ponti et~al.(2004)Ponti, Machacek, Gupton, Waterman-Storer and
  Danuser}]{ponti2004two}
\bibinfo{author}{Ponti, A.}, \bibinfo{author}{Machacek, M.},
  \bibinfo{author}{Gupton, S.L.}, \bibinfo{author}{Waterman-Storer, C.M.},
  \bibinfo{author}{Danuser, G.}, \bibinfo{year}{2004}.
\newblock \bibinfo{title}{Two distinct actin networks drive the protrusion of
  migrating cells}.
\newblock \bibinfo{journal}{Science} \bibinfo{volume}{305},
  \bibinfo{pages}{1782--1786}.
\bibitem[{Proti{{\`e}}re et~al.(2006)Proti{{\`e}}re, Boudaoud and
  Couder}]{Protiere+Couder.2006.1}
\bibinfo{author}{Proti{{\`e}}re, S.}, \bibinfo{author}{Boudaoud, A.},
  \bibinfo{author}{Couder, Y.}, \bibinfo{year}{2006}.
\newblock \bibinfo{title}{Particle{{--}}wave association on a fluid interface}.
\newblock \bibinfo{journal}{J. Fluid Mech.} \bibinfo{volume}{554},
  \bibinfo{pages}{85}.
\newblock \DOIprefix\doi{10.1017/S0022112006009190}.
\bibitem[{Putelat et~al.(2018)Putelat, Recho and
  Truskinovsky}]{putelat2018mechanical}
\bibinfo{author}{Putelat, T.}, \bibinfo{author}{Recho, P.},
  \bibinfo{author}{Truskinovsky, L.}, \bibinfo{year}{2018}.
\newblock \bibinfo{title}{Mechanical stress as a regulator of cell motility}.
\newblock \bibinfo{journal}{Physical Review E} \bibinfo{volume}{97},
  \bibinfo{pages}{012410}.
\bibitem[{Recho et~al.(2013)Recho, Putelat and
  Truskinovsky}]{recho2013contraction}
\bibinfo{author}{Recho, P.}, \bibinfo{author}{Putelat, T.},
  \bibinfo{author}{Truskinovsky, L.}, \bibinfo{year}{2013}.
\newblock \bibinfo{title}{Contraction-driven cell motility}.
\newblock \bibinfo{journal}{Physical review letters} \bibinfo{volume}{111},
  \bibinfo{pages}{108102}.
\bibitem[{Recho and Truskinovsky(2013)}]{recho2013asymmetry}
\bibinfo{author}{Recho, P.}, \bibinfo{author}{Truskinovsky, L.},
  \bibinfo{year}{2013}.
\newblock \bibinfo{title}{Asymmetry between pushing and pulling for crawling
  cells}.
\newblock \bibinfo{journal}{Physical Review E} \bibinfo{volume}{87},
  \bibinfo{pages}{022720}.
\bibitem[{Recho and Truskinovsky(2016)}]{Recho-Truskinovsky.2015.1}
\bibinfo{author}{Recho, P.}, \bibinfo{author}{Truskinovsky, L.},
  \bibinfo{year}{2016}.
\newblock \bibinfo{title}{Maximum velocity of self-propulsion for an active
  segment}.
\newblock \bibinfo{journal}{Math. Mech. Solids} \bibinfo{volume}{21},
  \bibinfo{pages}{263--278}.
\bibitem[{Ron et~al.(2020)Ron, Monzo, Gauthier, Voituriez and
  Gov}]{Ron+Gov.2020.1}
\bibinfo{author}{Ron, J.E.}, \bibinfo{author}{Monzo, P.},
  \bibinfo{author}{Gauthier, N.C.}, \bibinfo{author}{Voituriez, R.},
  \bibinfo{author}{Gov, N.S.}, \bibinfo{year}{2020}.
\newblock \bibinfo{title}{One-dimensional cell motility patterns}.
\newblock \bibinfo{journal}{Phys. Rev. Research} \bibinfo{volume}{2},
  \bibinfo{pages}{2435}.
\newblock \DOIprefix\doi{10.1103/PhysRevResearch.2.033237}.
\bibitem[{Rubinstein et~al.(2009)Rubinstein, Fournier, Jacobson, Verkhovsky and
  Mogilner}]{rubinstein2009actin}
\bibinfo{author}{Rubinstein, B.}, \bibinfo{author}{Fournier, M.F.},
  \bibinfo{author}{Jacobson, K.}, \bibinfo{author}{Verkhovsky, A.B.},
  \bibinfo{author}{Mogilner, A.}, \bibinfo{year}{2009}.
\newblock \bibinfo{title}{Actin-myosin viscoelastic flow in the keratocyte
  lamellipod}.
\newblock \bibinfo{journal}{Biophysical journal} \bibinfo{volume}{97},
  \bibinfo{pages}{1853--1863}.
\bibitem[{Sabass and Schwarz(2010)}]{sabass2010modeling}
\bibinfo{author}{Sabass, B.}, \bibinfo{author}{Schwarz, U.S.},
  \bibinfo{year}{2010}.
\newblock \bibinfo{title}{Modeling cytoskeletal flow over adhesion sites:
  competition between stochastic bond dynamics and intracellular relaxation}.
\newblock \bibinfo{journal}{Journal of Physics: Condensed Matter}
  \bibinfo{volume}{22}, \bibinfo{pages}{194112}.
\bibitem[{Schreiber et~al.(2021)Schreiber, Amiri, Heyn, R{\"a}dler and
  Falcke}]{schreiber2021adhesion}
\bibinfo{author}{Schreiber, C.}, \bibinfo{author}{Amiri, B.},
  \bibinfo{author}{Heyn, J.C.}, \bibinfo{author}{R{\"a}dler, J.O.},
  \bibinfo{author}{Falcke, M.}, \bibinfo{year}{2021}.
\newblock \bibinfo{title}{On the adhesion--velocity relation and length
  adaptation of motile cells on stepped fibronectin lanes}.
\newblock \bibinfo{journal}{Proceedings of the National Academy of Sciences}
  \bibinfo{volume}{118}, \bibinfo{pages}{e2009959118}.
\bibitem[{Sens(2013)}]{sens2013rigidity}
\bibinfo{author}{Sens, P.}, \bibinfo{year}{2013}.
\newblock \bibinfo{title}{Rigidity sensing by stochastic sliding friction}.
\newblock \bibinfo{journal}{EPL (Europhysics Letters)} \bibinfo{volume}{104},
  \bibinfo{pages}{38003}.
\bibitem[{Sens(2020)}]{Sens.2020.1}
\bibinfo{author}{Sens, P.}, \bibinfo{year}{2020}.
\newblock \bibinfo{title}{Stick{--}slip model for actin-driven cell
  protrusions, cell polarization, and crawling}.
\newblock \bibinfo{journal}{Proc Natl Acad Sci USA} \bibinfo{volume}{117},
  \bibinfo{pages}{24670--24678}.
\newblock \DOIprefix\doi{10.1073/pnas.2011785117}.
\bibitem[{Shenoy et~al.(2016)Shenoy, Wang and Wang}]{shenoy2016chemo}
\bibinfo{author}{Shenoy, V.B.}, \bibinfo{author}{Wang, H.},
  \bibinfo{author}{Wang, X.}, \bibinfo{year}{2016}.
\newblock \bibinfo{title}{A chemo-mechanical free-energy-based approach to
  model durotaxis and extracellular stiffness-dependent contraction and
  polarization of cells}.
\newblock \bibinfo{journal}{Interface focus} \bibinfo{volume}{6},
  \bibinfo{pages}{20150067}.
\bibitem[{Tjhung et~al.(2012)Tjhung, Marenduzzo and
  Cates}]{tjhung2012spontaneous}
\bibinfo{author}{Tjhung, E.}, \bibinfo{author}{Marenduzzo, D.},
  \bibinfo{author}{Cates, M.E.}, \bibinfo{year}{2012}.
\newblock \bibinfo{title}{Spontaneous symmetry breaking in active droplets
  provides a generic route to motility}.
\newblock \bibinfo{journal}{Proceedings of the National Academy of Sciences}
  \bibinfo{volume}{109}, \bibinfo{pages}{12381--12386}.
\bibitem[{Verkhovsky et~al.(1999)Verkhovsky, Svitkina and
  Borisy}]{verkhovsky1999self}
\bibinfo{author}{Verkhovsky, A.B.}, \bibinfo{author}{Svitkina, T.M.},
  \bibinfo{author}{Borisy, G.G.}, \bibinfo{year}{1999}.
\newblock \bibinfo{title}{Self-polarization and directional motility of
  cytoplasm}.
\newblock \bibinfo{journal}{Current Biology} \bibinfo{volume}{9},
  \bibinfo{pages}{11--S1}.
\bibitem[{Yam et~al.(2007)Yam, Wilson, Ji, Hebert, Barnhart, Dye, Wiseman,
  Danuser and Theriot}]{yam2007actin}
\bibinfo{author}{Yam, P.T.}, \bibinfo{author}{Wilson, C.A.},
  \bibinfo{author}{Ji, L.}, \bibinfo{author}{Hebert, B.},
  \bibinfo{author}{Barnhart, E.L.}, \bibinfo{author}{Dye, N.A.},
  \bibinfo{author}{Wiseman, P.W.}, \bibinfo{author}{Danuser, G.},
  \bibinfo{author}{Theriot, J.A.}, \bibinfo{year}{2007}.
\newblock \bibinfo{title}{Actin--myosin network reorganization breaks symmetry
  at the cell rear to spontaneously initiate polarized cell motility}.
\newblock \bibinfo{journal}{The Journal of cell biology} \bibinfo{volume}{178},
  \bibinfo{pages}{1207--1221}.
\bibitem[{Zhang et~al.(2020)Zhang, Rosakis, Hou and
  Ravichandran}]{zhang2020minimal}
\bibinfo{author}{Zhang, Z.}, \bibinfo{author}{Rosakis, P.},
  \bibinfo{author}{Hou, T.Y.}, \bibinfo{author}{Ravichandran, G.},
  \bibinfo{year}{2020}.
\newblock \bibinfo{title}{A minimal mechanosensing model predicts keratocyte
  evolution on flexible substrates}.
\newblock \bibinfo{journal}{Journal of the Royal Society Interface}
  \bibinfo{volume}{17}, \bibinfo{pages}{20200175}.

\end{thebibliography}

\end{document}